\documentclass[prd,twocolumn,superscriptaddress,floatfix,nofootinbib]{revtex4-2}
\pdfoutput=1 
\usepackage[T1]{fontenc} 
\usepackage{amsmath,amssymb,amsfonts,amsthm}
\usepackage{graphicx}
\usepackage[usenames,dvipsnames]{color}
\usepackage{enumitem}
\usepackage{verbatim}
\usepackage[percent]{overpic}
\usepackage{rotating}
\usepackage{hyperref}
\usepackage{array}
\usepackage{mathtools}
\usepackage{lmodern}
\usepackage[normalem]{ulem}
\usepackage{braket}
\usepackage{tensor}
\usepackage{dsfont}
\usepackage{bm}
\usepackage{tikz}

\usepackage{mathrsfs}

\usetikzlibrary{decorations.pathmorphing}
\usepackage{CJKutf8}

\usetikzlibrary{positioning}
\usetikzlibrary{calc,through,backgrounds}

\theoremstyle{definition}
\newtheorem{definition}{Definition}[section]
\newtheorem{theorem}{Theorem}[section]

\newcommand{\sgn}{\text{sgn}}
\newcommand{\kako}[1]{\left( #1 \right)}
\newcommand{\kagikako}[1]{\left[ #1 \right]}
\newcommand{\ts}[1]{ _{\text{#1}} }
\newcommand{\Bigkako}[1]{\Big( #1 \Big)}

\allowdisplaybreaks

\DeclareMathOperator{\Tr}{Tr}
\newcommand{\R}{\mathbb{R}}
\newcommand{\C}{\mathbb{C}}
\newcommand{\dd}{\text{d}}

\newcommand{\id}{\mathds{1}}

\newcommand{\ii}{\mathsf{i}}

\begin{document}

\title{Exact treatment of the memory kernel under time-dependent system-environment coupling via a train of delta distributions}

\author{Yuta Uenaga}
\email{uenaga.yuta.617@s.kyushu-u.ac.jp}
\affiliation{Department of Physics, Kyushu University, 
744 Motooka, Nishi-Ku, Fukuoka 819-0395, Japan}

\author{Kensuke Gallock-Yoshimura}
\email{gallockyoshimura@biom.t.u-tokyo.ac.jp} 

\affiliation{Department of Information and Communication Engineering, 
Graduate School of Information Science and Technology,
The University of Tokyo, 
7–3–1 Hongo, Bunkyo-ku, Tokyo 113–8656, Japan}
\affiliation{Department of Physics, Kyushu University, 
744 Motooka, Nishi-Ku, Fukuoka 819-0395, Japan}

\author{Takano Taira}
\email{taira.takano.292@m.kyushu-u.ac.jp} 

\affiliation{Department of Physics, Kyushu University, 
744 Motooka, Nishi-Ku, Fukuoka 819-0395, Japan}
\affiliation{Institute of Industrial Science, The University of Tokyo, 5-1-5 Kashiwanoha, Kashiwa, Chiba 277-8574, Japan}

\begin{abstract}
Memory effects in a quantum system coupled to an environment are one of the central features in the theory of open quantum systems. 
The dynamics of such quantum systems are typically governed by an equation of motion with a time-convolution integral of the memory kernel. 
However, solving such integro-differential equations is challenging, especially when the memory kernel is nonstationary (not time-translation invariant). 
In this paper, we analytically and nonperturbatively solve such integro-differential equations with a nonstationary memory kernel by employing a train of Dirac-delta switchings. 
We then apply this method to the damped Jaynes-Cummings model and the damped harmonic oscillator model to demonstrate that (i) our solution asymptotes to the well-known exact solution in the continuum limit, and that (ii) our method also enables us to visualize the memory effect in the environment. 
\end{abstract}

\maketitle
\flushbottom

\section{Introduction}
The dissipative behavior of a quantum system interacting with its environment is one of the central features of open quantum systems \cite{weiss2012quantum, breuer2002theory}. 
One of the key aspects in the study of open quantum systems is the memory effect in the environment. 
A Markovian dynamics is memoryless, which leads to one-way flow of information from a system to an environment. 
In contrast, a non-Markovian dynamics allows for the memory from the past interactions, which leads to a backflow of information from the environment to the system of interest \cite{Vega.nonMarkov.2017}.

Mathematically, such memory effects appear in equations of motion and cause difficulties. 
We are particularly interested in a quantity called the \textit{memory kernel} $\Sigma(t,t')$, which appears as a time-convolution integral (``an integral over the past history of the system'' \cite{breuer2002theory}): 
\begin{align}
    \int_{t_0}^t \dd t'\,
    \Sigma(t,t') \mathcal O(t')\,, \label{eq:time-convolution integral}
\end{align}
where $\mathcal O(t)$ is some observable. 
This structure arises, for example, in the Nakajima-Zwanzig equation \cite{nakajima1958quantum, Zwanzig.1960}---an exact integro-differential equation for the density matrix of the system of interest. 
Because such integro-differential equations are difficult to solve in general, analyses often rely on perturbation theory.

There are, however, special models that admit exact solutions even in the presence of a time-convolution integral. 
One example is a two-level system immersed in a bosonic bath. 
In particular, the damped Jaynes--Cummings model describes the coupling of a two-level system to a single-mode bosonic bath with a Lorentz spectral density. 
Another example is a damped harmonic oscillator obeying the Heisenberg equation of motion known as the quantum Langevin equation (QLE) \cite{weiss2012quantum, Ford.Statistical.1965, Ford.QLE.1988}. 
Specifically, the

Caldeira--Leggett model \cite{Caldeira.Leggett.Influence.1981, CALDEIRA.path.integral.1983, CALDEIRA.tunnelling.1983} is a cornerstone in the theory of open quantum systems, describing the dynamics of a quantum Brownian particle \cite{breuer2002theory}. 
In this framework, a quantum harmonic oscillator serves as the system, and it is coupled to an environment composed of a collection of quantum harmonic oscillators. 

While these exactly solvable models are extremely useful, they rely on the stationarity of the memory kernel.
In many studies, the focus is on the asymptotic behavior of the system, e.g., the density matrix or the two-time correlation functions of the system quadratures in the long-interaction limit, during which the system reaches equilibrium with the environment. 
Therefore, it is assumed that the system \textit{constantly} couples to the bath for an infinitely long time. 
In fact, this time-independent coupling is crucial for solving the master equation and QLE analytically, as the memory kernel in these models is time-translation invariant (i.e., stationary), $\Sigma(t,t')=\Sigma(t-t')$, and so the time-convolution integral \eqref{eq:time-convolution integral} can be treated without approximations. 
However, once we introduce a time-dependent coupling, the memory kernel becomes nonstationary, which makes it challenging to solve the equation of motion analytically. 
This issue is particularly relevant in the study of finite-time interactions in, e.g., quantum thermodynamics \cite{CANGEMI.review.2024}.

In this paper, we propose a nonperturbative method for solving equations of motion with the time-convolution integral, even when the memory kernel is nonstationary. 
In particular, we consider a class of integro-differential equations that contains the memory kernel of the form 
\begin{align}
    \Sigma(t,t') = \chi(t) \chi(t') \Gamma(t,t')\,, \label{eq:memory kernel with switching}
\end{align}
where $\chi(t)$ is the so-called switching function, which governs the time-dependence of the interaction, and $\Gamma(t,t')$ is a function characterized by the spectral density of the environment. 
We then employ a train of Dirac-delta couplings (a sum of Dirac's delta distributions) to mimic a continuous time-dependent coupling, allowing us to solve the equations of motion analytically. 
This method is equivalent to Refs.~\cite{Deramon.spinboson.2020, Jose.train.delta.2024}, where a train of Delta switchings is employed to nonperturbatively analyze the behavior of qubits coupled to a quantum scalar field in curved spacetime. 
In particular, Ref.~\cite{Jose.train.delta.2024} mathematically showed that, in the continuum limit, the density matrix obtained by the train-of-delta method converges to the original density matrix with a continuous switching function. 
In this paper, we show that this method simplifies the handling of the memory kernel, the main barrier to solving the integro-differential equation. 
Although one might initially think that a time-local delta distribution only produces the Markovian dynamics, we show that, for the Jaynes--Cummings and the Caldeira--Leggett models, a collection of delta distributions can remarkably capture the non-Markovian behavior of the environment. 
In fact, the train of Dirac delta switchings enables us to pictorially understand how the memory effect in the environment plays a role. 
We show how to construct the diagram depicting the environment's memory effect.

This paper is organized as follows. 
In Sec.~\ref{subsec:solving integro-diff eq}, we consider a general inhomogeneous integro-differential equation \eqref{eq:general integro-diff eq} and analytically solve it by employing a train of delta-switchings \eqref{eq:train of delta}. 
The main result is given by Eq.~\eqref{eq:main solution}. 
The method we use allows us to visualize the memory effect using diagrams, which will be explained in Sec.~\ref{subsec:pictorial}. 
We then apply our main result \eqref{eq:main solution} to two concrete models: the damped Jaynes-Cummings model (Sec.~\ref{sec:JC model}) and the damped harmonic oscillator model (Sec.~\ref{sec:QLE}). 
In a simple example of the damped Jaynes-Cummings model, we compare our solution to the well-known exact solution in Sec.~\ref{subsec:JC solving QME}, and examine the relationship between our diagrams and the non-Markovianity in Sec.~\ref{subsec:nonmarkov JC model}.  
We also analyze the QLE for the damped harmonic oscillator model in Sec.~\ref{sec:QLE}. 
After applying our method to the QLE in Sec.~\ref{subsec:Solving QLE}, we consider the two-time correlation functions and the covariance matrix of the system observables in Sec.~\ref{subsec:correlation function and CM}. 
We then choose the Lorentz-Drude spectral density in Sec.~\ref{subsec:spectral density} and compare our solution to the exact solution in Sec.~\ref{subsec:long interaction}. 

Throughout this paper, we set $\hbar = c=k\ts{B}=1$ and use the convention where the system's mass is $M=1$.

\section{Method: a train of Dirac's delta distributions}\label{sec:method}
\subsection{Solving an integro-differential equation}\label{subsec:solving integro-diff eq}

To begin with, we explain our main idea for solving an integro-differential equation by using a train of Dirac's delta distributions. 
Consider an arbitrary system-environment coupling described by the following total Hamiltonian: 
\begin{align}
    H\ts{tot}(t)=
        H\ts{s,0} 
        + 
        H\ts{E,0}
        +
        H\ts{int}(t)\,,
\end{align}
where $H\ts{s,0}$ and $H\ts{E,0}$ are the free Hamiltonians of the system and environment, respectively. 
The interaction Hamiltonian in the Schr\"odinger picture, $H\ts{int}(t)$, is generically expressed as 
\begin{align}
    H\ts{int}(t)
    =
        \sum_{\alpha} c_\alpha(t) \mathcal O_\alpha \otimes B_\alpha\,,
\end{align}
where $c_\alpha(t) (\geq 0)$ is the time-dependent coupling constant, $\{ \mathcal O_\alpha\}_\alpha$ and $\{ B_\alpha \}_\alpha$ are the collections of operators of the system and environment, respectively. 
For convenience, we separate the function characterizing the time-dependence of the interaction from $c_\alpha(t)$ by writing 
\begin{align}
    c_\alpha(t)=c_\alpha \chi(t)\,.
\end{align}
The continuous function $\chi(t)$ is referred to as the \textit{switching function}. 
In what follows, we assume that the interaction begins at time $t=0$.

When no interaction is involved (i.e., $c_\alpha=0$ for all $\alpha$), the system of our interest freely evolves under the following linear ordinary differential equation of order $n$: 
\begin{align}
    p(\partial_t)\mathcal O(t)=0\,,
\end{align}
where $\partial_t \equiv \frac{\dd}{\dd t}$, and $p(\partial_t)$ is a polynomial differential operator of the form 
\begin{align}
    p(\partial_t)
    \equiv 
        \sum_{k=0}^n a_k \dfrac{\dd^k}{\dd t^k}\,.
\end{align}
For example, the Liouville-von Neumann equation for the system's density matrix $\rho\ts{s}(t)$ in the Schr\"odinger picture is 
\begin{align}
    \dfrac{\dd }{\dd t}\rho\ts{s}(t)
    + 
    \ii [H\ts{s,0}, \rho\ts{s}(t)]=0\,. \notag 
\end{align}
Another example is the Heisenberg equation of motion. 
In particular, the Heisenberg equation of motion for a free quantum harmonic oscillator reads 
\begin{align}
    \dfrac{\dd^2}{\dd t^2} Q(t) + \Omega^2 Q(t)=0\,, \notag 
\end{align}
where $\Omega$ is the frequency. 
In this case, we have $p(\partial_t)=\partial_t^2 + \Omega^2$. 
In this paper, we focus on $n=1$ and 2.

We now consider the interaction between such a quantum system and an environment. 
Specifically, we consider a class of integro-differential equations of the form 
\begin{align}
    p(\partial_t) \mathcal O(t)
    + 
    \int_0^t \dd t'\, \chi(t) \chi(t') \Gamma(t,t') \mathcal O(t')
    =
    \chi(t)\xi(t)\,, \label{eq:general integro-diff eq}
\end{align}
where $\chi(t)$ is a switching function that controls the time-dependence of coupling, and $\xi(t)$ is an inhomogeneous term, typically known as the noise term. 
The second term on the left-hand side is a time-convolution integral with the memory kernel $\Sigma(t,t')\equiv \chi(t) \chi(t') \Gamma(t,t')$, which is nonstationary unless the coupling is time-independent ($\chi(t)=1$) and the environment is stationary, $\Gamma(t,t')=\Gamma(t-t')$. 
The quantity $\Gamma(t,t')$ represents how information flows through the environment, and we assume that the memory effect does not influence the system instantaneously: 
\begin{align}
    t=t' \Rightarrow \Gamma(t,t')=0\,. \label{eq:zero insta memory}
\end{align}
The memory kernel contains two switching functions because $\Sigma(t,t')$ is a two-time correlation function of the form $c_\alpha(t)c_{\alpha'}(t') \Tr[ B_\alpha (t) B_{\alpha'}(t')  ]$.

The integro-differential equation \eqref{eq:general integro-diff eq} is known to be analytically solvable when the memory kernel is stationary. 
Let us set $\chi(t)=1$ and assume $\Gamma(t,t')=\Gamma(t-t')$. 
We then perform the Laplace transformation $\cal L$ defined by
\begin{align}
    \tilde f(z)
    \equiv 
    \mathcal L[f(t)]
    \coloneqq
        \int_0^\infty \dd t \, f(t) e^{-zt}\,.
        \quad z\in \C
\end{align}
Then Eq.~\eqref{eq:general integro-diff eq} becomes 
\begin{align}
    \tilde{\mathcal O}(z)
    =
        \tilde{G}(z)
        \sum_{k=1}^n a_k
        \sum_{j=0}^{k-1} z^{k-j-1} \left.\dfrac{\dd^j \mathcal O(t)}{\dd t^j}\right|_{t=0}
        +
        \tilde{G}(z) \tilde \xi(z)\,,
\end{align}
where $\tilde G(z)\coloneqq [p(z)+\tilde \Gamma(z)]^{-1} $ with $p(z)=\sum_{k=0}^n a_k z^k$, and we have used the following properties: 
\begin{align}
    &\mathcal L
    \kagikako{
        \dfrac{\dd^n \mathcal O(t)}{\dd t^n}
    }
    =
        z^n \tilde{\mathcal O}(z)
        - \left. \sum_{j=0}^{n-1} z^{n-j-1} \dfrac{\dd^j \mathcal O(t)}{\dd t^j}\right|_{t=0}\,, \notag \\
    &\mathcal L
    \int_0^t \dd t'\, \Gamma(t-t')\mathcal O(t')
    =
        \tilde \Gamma(z) \tilde{\mathcal O}(z)\,. \label{eq:convoluted Laplace property}
\end{align}
The solution to the integro-differential equation \eqref{eq:general integro-diff eq} is obtained by performing the inverse Laplace transform: 
\begin{align}
    \mathcal O(t)
    &=
        \sum_{k=1}^n a_k
        \sum_{j=0}^{k-1} 
        \mathcal{L}^{-1}[\tilde{G}(z) z^{k-j-1} ]
        \left.\dfrac{\dd^j \mathcal O(t)}{\dd t^j}\right|_{t=0} \notag \\
        &\quad
        +
        \int_0^t \dd t'\,G(t-t') \xi(t')\,,
\end{align}
where $G(t)\equiv \mathcal L^{-1}[\tilde G(z)]$ is the Green function. 
For example, if we are considering a second-order differential equation ($n=2$), then we have 
\begin{align}
    \mathcal O(t)
    &=
        a_2 
        \kagikako{
            \dot G(t) \mathcal O(0) 
            +
            G(t) \dot{\mathcal{O}}(0)
        }
        +
        a_1 G(t) \mathcal O(0) \notag \\
        &\quad
        + 
        \int_0^t \dd t'\,G(t-t') \xi(t')\,,
\end{align}
where $\dot{\mathcal O}(0)\equiv \left.\frac{\dd \mathcal O(t)}{\dd t}\right|_{t=0}$. 
From the derivation above, we see that the stationary memory kernel $\Sigma(t,t')=\Gamma(t-t')$ allows us to simplify the time-convolution integral by using the property \eqref{eq:convoluted Laplace property}. 
However, this is no longer applicable if the memory kernel is nonstationary.

We now present our main idea for solving the integro-differential equation \eqref{eq:general integro-diff eq} when the memory kernel is nonstationary. 
Suppose the switching function is compactly supported on $t\in [0,T]$. 
We choose the switching function in such a way that it is represented by $N$-Dirac delta distributions \cite{Deramon.spinboson.2020, Jose.train.delta.2024}:  
\begin{align}
    \chi(t)
    &=
        \dfrac{T}{N} 
        \sum_{k=1}^N 
        \chi
        \kako{
            \dfrac{k}{N} T
        }
        \delta 
        \kako{
            t - \dfrac{k}{N}T
        } \notag \\
    &\equiv 
        \dfrac{T}{N} 
        \sum_{k=1}^N 
        \chi (t_k) \delta (t-t_k)\,. \label{eq:train of delta}
\end{align}
Here, the Dirac deltas are uniformly distributed between $t\in [0,T]$, and the $k$-th delta coupling occurs at time $t_k \equiv kT/N$ with the amplitude $\chi(t_k)$. 
Note that $T/N$ is the interval between successive interactions, $t_{k}$ and $t_{k+1}$. 
This allows us to simplify the time-convolution integral \eqref{eq:time-convolution integral}. 
Specifically, the Laplace transform of this integral reads 
\begin{align}
    &\mathcal L
    \int_0^t \dd t'\,
    \Sigma(t,t') \mathcal O(t') \notag \\
    &=
        \mathcal L
        \int_0^\infty \dd t'\,
        \chi(t)\chi(t')\Gamma(t,t') \mathcal O(t') \Theta(t-t') \notag \\
    &=
        \dfrac{T^2}{N^2}
        \sum_{k,l=1}^N 
        \Sigma(t_k,t_l)
        \Theta(t_k - t_l)
        \mathcal O(t_l) e^{ -z t_k }\,,
\end{align}
where $\Sigma(t_k, t_l)\equiv \chi(t_k) \chi(t_l) \Gamma(t_k,t_l)$ and $\Theta(t)$ is the Heaviside step function with the convention $\Theta(0)=1$. 
The noise term $\zeta(t)\equiv\chi(t)\xi(t)$ can also be transformed as 
\begin{align}
    \mathcal{L}[ \zeta(t) ] 
    &=
        \sum_{k=1}^N 
        \zeta(t_k) e^{ -z t_k }\,,
    \quad
    \zeta(t_k)
    \equiv 
        \dfrac{T}{N} \chi(t_k) \xi(t_k)\,. \label{eq:Laplace of zeta}
\end{align}
Therefore, after applying the Laplace transformation to the integro-differential equation \eqref{eq:general integro-diff eq} and obtaining
\begin{align}
    \tilde{\mathcal O}(z)
    &=
        \sum_{k=1}^n a_k 
        \sum_{j=0}^{k-1} \dfrac{z^{k-j-1}}{p(z)} 
        \left. \dfrac{\dd^j \mathcal O(t)}{\dd t^j}\right|_{t=0} \notag \\
        &\quad
        -\dfrac{T^2}{N^2}
        \sum_{k,l=1}^N 
        \Sigma(t_k, t_l) 
        \Theta(t_k-t_l)
        \dfrac{e^{-z t_k}}{ p(z) } 
        \mathcal{O}(t_l) \notag \\
        &\quad
        +
        \sum_{k=1}^N \zeta(t_k) \dfrac{e^{-z t_k}}{ p(z) } \,,
\end{align}
we perform the inverse Laplace transformation to obtain
\begin{align}
    \mathcal{O}(t)
    &=
         \mathcal{O}_t^{(0)}
         +
         \sum_{l=1}^N K_{t,t_l} \mathcal{O}_l
         + 
         \Xi_t\,, \label{eq:Q(t) result intermediate}
\end{align}
where 
\begin{subequations}
\begin{align}
    &\mathcal{O}_t^{(0)}
    \coloneqq
        \sum_{k=1}^n a_k 
        \sum_{j=0}^{k-1} 
        \mathcal{L}^{-1}
        \kagikako{
            \dfrac{z^{k-j-1}}{p(z)} 
        }
        \left. \dfrac{\dd^j \mathcal O(t)}{\dd t^j}\right|_{t=0} \,, \\
    &K_{t, t_l}
    \equiv 
        -\dfrac{T^2}{N^2}
        \sum_{k=1}^N 
        G_0(t-t_k) \Theta(t-t_k)
        \Sigma(t_k,t_l)
        \Theta(t_k-t_l)  \,, \label{eq:Ktl} \\
    &\mathcal{O}_l
    \equiv 
        \mathcal{O}(t=t_l)\,, \\
    &\Xi_t
    \equiv 
        \sum_{k=1}^N 
        G_0(t-t_k) \Theta(t-t_k)
        \zeta(t_k) \,. \label{eq:Xit}
\end{align}
\end{subequations}
Here, $Q_t^{(0)}$ is the solution to the differential equation for the free theory, $p(\partial_t)\mathcal O(t)=0$, which can be solved once the initial conditions are specified. 
$K_{t, t_l}$ is the term that emerges from the memory kernel (thus, responsible for the memory effect) with the property $K_{t,t_N}=0$ due to $\Sigma(t_N,t_N)=0$ [Eq.~\eqref{eq:zero insta memory}]. 
In $K_{t,t_l}$, we have the Green function $G_0(t) \equiv \mathcal L^{-1}[1/p(z)]$ of the differential equation without the memory kernel, i.e., it satisfies $p(\partial_t) G_0(t)=\delta(t)$. 
We also used the property
\begin{align}
    \mathcal{L}^{-1}
        \kagikako{
            \dfrac{e^{-z t_k}}{ p(z) } 
        }
    =
        G_0(t-t_k) \Theta(t-t_k)\,. \notag 
\end{align}
Thus, the term $G_0(t-t_k) \Theta(t-t_k)$ represents how $\mathcal O(t)$ reacts to a delta-input at time $t_k$ when the memory kernel is absent. 
From this observation, the quantity $K_{t,t_l}$ with a fixed $t_l$ can be interpreted as follows: 
a delta coupling at time $t_l$ affects the system-environment coupling at time $t_k$ in the future through the memory kernel $\Sigma(t_k, t_l)$, and this effect freely propagates to time $t(>t_k)$ through $G_0$. 
By summing over all possible propagations $t_l \xrightarrow{\Sigma} t_k \xrightarrow{G_0} t$, we obtain $K_{t,t_l}$.

Although $\mathcal{O}(t)$ in \eqref{eq:Q(t) result intermediate} takes a very simple form, it still depends on itself in the past, $\mathcal{O}_l$. 
Below, we obtain the general solution $\mathcal{O}(t)$ that depends only on the initial values $\frac{\dd^j \mathcal{O}}{\dd t^j}|_{t=0}$. 
Substituting $t=t_l$ in \eqref{eq:Q(t) result intermediate} gives us 
\begin{align}
    \mathcal{O}_l 
    &=
        \mathcal{O}_l^{(0)}
        + 
        \sum_{i=1}^N K_{li} \mathcal{O}_i
        + 
        \Xi_l\,,
\end{align}
where $\mathcal{O}_l^{(0)}$ and $\Xi_l$ are understood as $\mathcal{O}_l^{(0)}\equiv \mathcal{O}^{(0)}(t_l)$ and $\Xi_l\equiv \Xi(t_l)$. 
The equation above can be written in terms of vectors and matrices as follows: 
\begin{align}
    \bm{\mathcal{O}}
    &=
        \bm{\mathcal{O}}^{(0)}
        + \mathbf{K} \bm{\mathcal{O}}
        + \bm \Xi\,,  \label{eq:Q matrix eq}
\end{align}
where 
\begin{subequations}
\begin{align}
    \bm{\mathcal{O}}
    &\coloneqq
        [ \mathcal{O}_1, \mathcal{O}_2, \ldots, \mathcal{O}_N ]^\intercal\,, \\
    \bm{\mathcal{O}}^{(0)}
    &\coloneqq
        [ \mathcal{O}^{(0)}_1, \mathcal{O}^{(0)}_2, \ldots, \mathcal{O}^{(0)}_N ]^\intercal\,, \\
    \mathbf{K}
    &\coloneqq
        \begin{bmatrix}
            0 & 0 & 0 & \ldots & 0\\
            K_{21} & 0 & 0 & \ldots & 0 \\
            K_{31} & K_{32} & 0 & \ldots  & 0 \\
            \vdots & \vdots & \ddots & \ddots & \vdots \\
            K_{N1} & K_{N2} & \ldots & K_{N,N-1} & 0 \\
        \end{bmatrix}\,, \\
    \bm \Xi
    &\coloneqq
        [ \Xi_1, \Xi_2, \ldots, \Xi_N ]^\intercal\,. 
\end{align}
\end{subequations}
Here, $K_{li}$ is the elements of the $N\times N$ matrix $\mathbf{K}$ given by 
\begin{align}
    K_{l i}
    &=
        -\dfrac{T^2}{N^2}
        \sum_{k=1}^N 
        G_0(t_l-t_k)
        \Theta(t_l-t_k)
        \Sigma(t_k,t_i)
        \Theta(t_k-t_i) 
         \,, \label{eq:Kli}
\end{align}
and the matrix $\mathbf{K}$ is a strictly lower triangular matrix due to  Heaviside's step functions in $K_{li}$.

The equation \eqref{eq:Q matrix eq} can be solved for $\bm{\mathcal{O}}$ as 
\begin{align}
    \bm{\mathcal{O}}
    &=
        (\mathbf{I} - \mathbf{K})^{-1} \bm{\mathcal{O}}^{(0)}
        +
        (\mathbf{I} - \mathbf{K})^{-1} \bm \Xi\,,
\end{align}
where $\mathbf{I}$ is the $N\times N$ identity matrix. 
Furthermore, the $N\times N$ strictly lower triangular matrix $\mathbf{K}$ has the nilpotent property: $\mathbf{K}^N=\mathbf{O}$, where $\mathbf{O}$ is the zero matrix. 
This allows us to express the inverse matrix $(\mathbf{I} - \mathbf{K})^{-1}$ as\footnote{Here, $\mathbf{K}^0 \equiv \mathbf{I}$.} 
\begin{align}
    (\mathbf{I} - \mathbf{K})^{-1}
    &=
        \sum_{n=0}^{N-1} \mathbf{K}^n 
    \equiv 
        \mathsf{K}\,, \label{eq:mathsfK}
\end{align}
because 
\begin{align}
    &(\mathbf{I} - \mathbf{K})
    (\mathbf{I} + \mathbf{K} + \mathbf{K}^2 + \ldots + \mathbf{K}^{N-1}) \notag \\
    &=
        \mathbf{I} - \mathbf{K}^N
    =
        \mathbf{I} - \mathbf{O}
    = \mathbf{I}\,.
\end{align}

We thus obtained a compact form of the vector of $\mathcal{O}_l$: 
\begin{align}
    \bm{\mathcal{O}}
    &=
        \mathsf{K} \bm{\mathcal{O}}^{(0)}
        +
        \mathsf{K} \bm \Xi\,. \label{eq:main result special case}
\end{align}
Substituting $\mathcal{O}_l$ into $\mathcal{O}(t)$ in \eqref{eq:Q(t) result intermediate}, we finally reached our main result: 
the solution to the integro-differential equation \eqref{eq:general integro-diff eq} with a train of delta distributions is given by 
\begin{align}
    \mathcal{O}(t)
    &=
        \mathcal{O}^{(0)}_t + \Xi_t
        +
        \sum_{l,i=1}^N K_{t,t_l} \mathsf{K}_{li}
        \Bigkako{
            \mathcal{O}^{(0)}_i + \Xi_i
        }\,. \label{eq:main solution}
\end{align}
The first term $\mathcal{O}^{(0)}_t$ is the free solution, and the second term $\Xi_t$ given by \eqref{eq:Xit} corresponds to the superposition of all possible noise impulses. 
The last term exhibits how the memory kernel affects the system. 
In the next section, we give the interpretation of such memory effects and visualize them using diagrams.

\subsection{Pictorial interpretation}\label{subsec:pictorial}
To understand the implication of the main result \eqref{eq:main solution}, we first note that $\mathcal O^{(0)}_i\equiv \mathcal O^{(0)}(t_i)$ is the value of the observable at time $t_i$ when freely evolved, and $\Xi_i \equiv \Xi(t_i)$ represents the effect of noise occurred during $t\in (0,t_i)$. 
The system then instantaneously couples to the environment at time $t_i$, followed by $K_{t,t_l} \mathsf{K}_{li}$. 
The elements of the matrix $\mathsf{K}$ in \eqref{eq:mathsfK} consist of $K_{li}$, and it reads 
\begin{align}
    \mathsf K_{jk}
    =
        \delta_{jk}
        +
        \sum_{n=0}^{N-2} 
        \sum_{j=i_0>i_1>\ldots>i_{n}>i_{n+1}=k}
        \prod_{r=0}^{n}
        K_{i_r i_{r+1}} \,.
\end{align} 
For instance, 
\begin{align}
    \mathsf{K}
    &=
        \begin{bmatrix}
            1 &0 \\
            K_{21} & 1
        \end{bmatrix}
    \quad
    \text{for }N=2\,, \notag \\
    \mathsf{K}
    &=
        \begin{bmatrix}
            1 & 0 & 0 \\
            K_{21} & 1 & 0 \\
            K_{31}+K_{32}K_{21} & K_{32} & 1
        \end{bmatrix}
    \quad
    \text{for }N=3\,. \notag 
\end{align}
As we explained in the previous subsection, $K_{t,t_i}$ in \eqref{eq:Ktl} corresponds to the memory propagation $t_i \xrightarrow{\Sigma} t_k \xrightarrow{G_0} t$. 
Similarly, $K_{li}$ can be interpreted as the propagation to $t=t_l$: $t_i \xrightarrow{\Sigma} t_k \xrightarrow{G_0} t_l$.
From the examples of $\mathsf K$ above, we see that the element $\mathsf{K}_{li}$ consists of all possible memory propagations from $t_i$ to $t_l$. 
Although such a term appears cumbersome, we show that it can be intuitively understood using diagrams.

\begin{figure}[t]
\centering
\includegraphics[width=\linewidth]{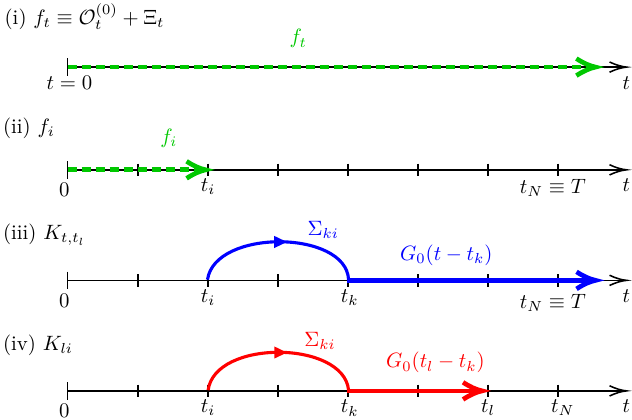}
\caption{The basic diagrams that constitute \eqref{eq:Gft}. 
Each $t_i$, $i\in \{ 1,2, \ldots, N \}$ represents the time at which the instantaneous interaction occurs. 
Here, we use a simplified notation $\Sigma_{ki}\equiv \Sigma(t_k, t_i)$. 
}
\label{fig:diagrams}
\end{figure}

To demonstrate, we express the solution \eqref{eq:main solution} as 
\begin{align}
    \mathcal O(t)
    &=
        f_t
        +
        \sum_{l,i=1}^N K_{t, t_l} \mathsf{K}_{li} f_i\,, \label{eq:Gft}
\end{align}
where $f_t\equiv \mathcal O^{(0)}_t + \Xi_t$. 
Let us introduce diagrams that characterize each term in \eqref{eq:Gft}. 
Since the function $f_t$ describes $\mathcal O(t)$ in the absence of the memory kernel, we depict $f_t$ as a dashed line along the time axis, as shown in Fig.~\ref{fig:diagrams}(i). 
In particular, $f_i\equiv f(t_i)$ is illustrated as a dashed line from $t=0$ to $t=t_i$ [Fig.~\ref{fig:diagrams}(ii)].

For the memory propagation $K_{t,t_l}$, defined in \eqref{eq:Ktl}, we assign a solid arc to $\Sigma(t_k, t_l)$ and a line to $G_0(t-t_k)\Theta(t - t_k)$ as illustrated in Fig.~\ref{fig:diagrams}(iii). 
A similar interpretation applies to $K_{li}$ defined in \eqref{eq:Kli}, except that the propagation terminates at $t=t_l$ [Fig.~\ref{fig:diagrams}(iv)]. 
We again note that $\Sigma(t_k, t_k)=0$ for any environment. 
This means that an ``instantaneous loop'', which is closed at a single time $t=t_k$, is not allowed. 
Physically, it guarantees that there is no immediate backreaction from the environment to the system.

\begin{figure*}[t]
\centering
\includegraphics[width=\linewidth]{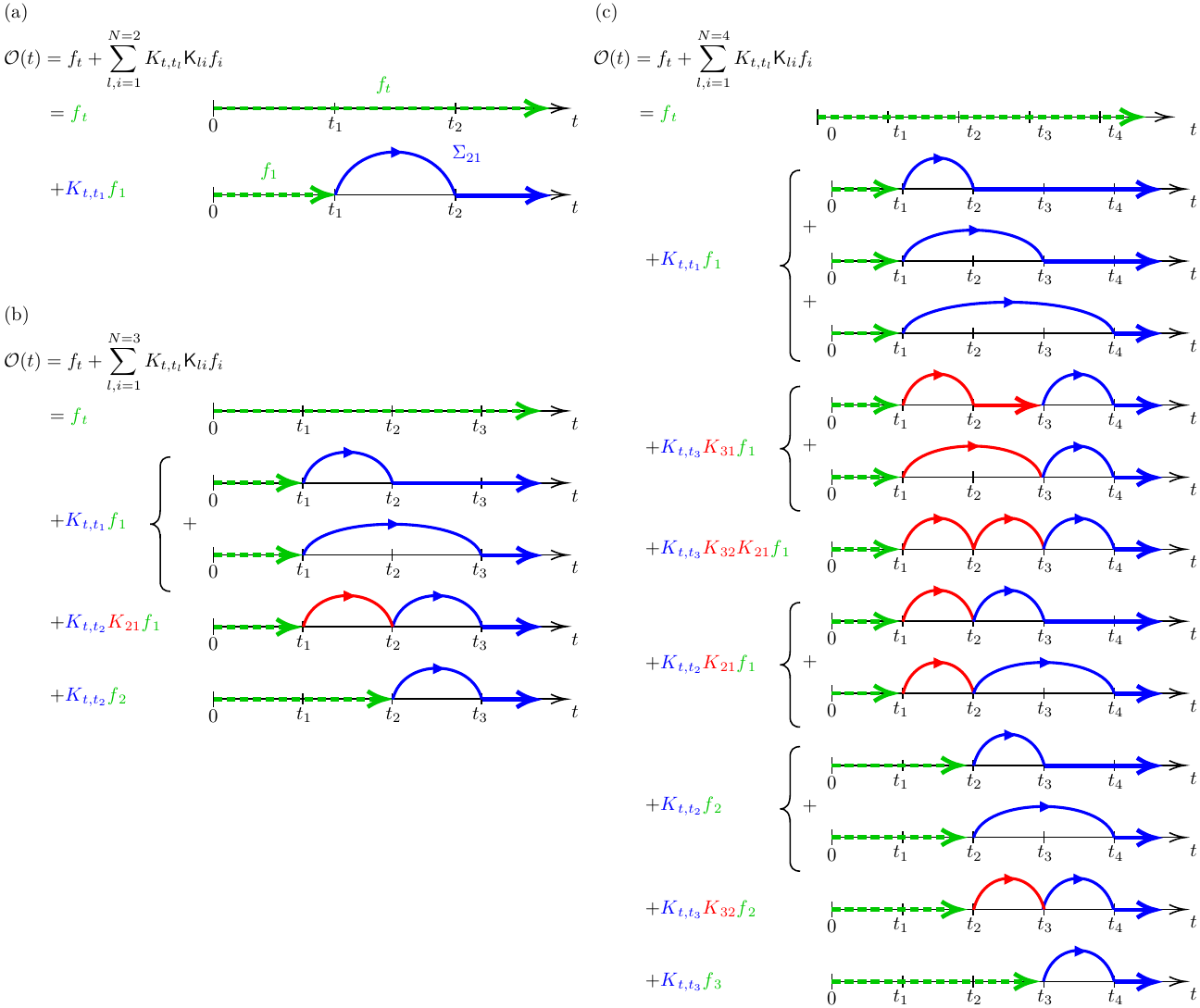}
\caption{
The diagrams illustrating $\mathcal O(t)$ in \eqref{eq:Gft} for (a) $N=2$, (b) $N=3$, and (c) $N=4$. 
An arc connecting $t_i$ and $t_j$ ($j>i$) corresponds to $\Sigma(t_j, t_i) \equiv \Sigma_{ji}$. 
}
\label{fig:N2diagram}
\end{figure*}

Let us take a simple example of $N=2$ and visualize $\mathcal O(t)$ in \eqref{eq:Gft}. 
Reminding that $K_{t,t_N}=0$, the solution $\mathcal O(t)$ for $N=2$ (delta-coupling twice) can be explicitly written as 
\begin{align}
    \mathcal O(t)
    &=
        f_t
        +
        \sum_{l,i=1}^{N=2} K_{t, t_l} \mathsf{K}_{li} f_i \notag \\
    &=
        f_t 
        + 
        \begin{bmatrix}
            K_{t,t_1} & 0
        \end{bmatrix}
        \begin{bmatrix}
            1 &0 \\
            K_{21} & 1
        \end{bmatrix}
        \begin{bmatrix}
            f_1 \\
            f_2
        \end{bmatrix} \notag \\
    &=
        f_t
        +
        K_{t,t_1} f_1 \,.
\end{align}
Figure~\ref{fig:N2diagram}(a) illustrates this expression using our diagrammatic elements. 
The term $K_{t,t_1} f_1$ can be understood as follows. 
First, draw a dashed line representing $f_1$ from $t=0$ to $t_1$. 
Then, we consider the factor $K_{t,t_1}$, which contains an arc and a line as illustrated in Fig.~\ref{fig:diagrams}(iii). 
In general, all possible configurations must be considered for each term. 
In the $N=2$ case, there is only one configuration for $K_{t,t_1}f_1$ as shown in Fig.~\ref{fig:N2diagram}(a). 
However, as we explain below, there will be multiple different configurations for a single term such like $K_{t,t_1}f_1$.

For $N=3$, we have 
\begin{align}
    \mathcal O(t)
    =
        f_t 
        + 
        K_{t,t_1}f_1
        +
        K_{t,t_2} K_{21} f_1
        +
        K_{t,t_2} f_2\,,
\end{align}
and the corresponding diagram is depicted in Fig.~\ref{fig:N2diagram}(b). 
Unlike the $N=2$ case, there are two possible configurations for $K_{t,t_1}f_1$, and their sum gives the final value of $K_{t,t_1}f_1$. 
We also have the term $K_{t,t_2}K_{21}f_1$, which contains two arcs arising from $K_{t,t_2}$ and $K_{21}$. 
Here, $K_{21}$ is represented by an arc starting from $t_1$ and a line ending at $t_2$. 
For $N=4$, various configurations emerge [Fig.~\ref{fig:N2diagram}(c)].

Overall, the expression for $\mathcal O(t)$ sums over all possible memory effects of the environment. 
We stress that both Markovian and non-Markovian effects can also be understood using the diagrams. 
Markovian processes correspond to diagrams that consist of arcs connecting adjacent times, namely, the terms with $\Sigma(t_{i+1}, t_i)$. 
One can be convinced by considering the ``continuum limit'' of the delta switchings, $N\to \infty$ and $t_{i+1}-t_i(\equiv T/N) \to 0$. 
In this case, the adjacent memory effect $\Sigma(t_{i+1},t_i)$ can be considered time-local. 
On the other hand, non-Markovian effects are represented by arcs connecting distant points $t_i$ and $t_{i+j}$ with $j\geq 2$. 
In Fig.~\ref{fig:N2diagram}(c), the Markovian diagrams correspond to $\Sigma_{21}$, $\Sigma_{32}$, and $\Sigma_{43}$. 
In fact, we will prove in Appendices~\ref{sec:Markov JC} and \ref{app:BornMarkov} that the solution $\mathcal O(t)$ composed only of adjacent arcs is Markovian.

In the next two sections (Secs.~\ref{sec:JC model} and \ref{sec:QLE}), we will apply our main result~\eqref{eq:main solution} to two concrete examples, namely the damped Jaynes–Cummings model and the Caldeira–Leggett model. 
For the Jaynes--Cummings model in Sec.~\ref{sec:JC model}, we will investigate the Markovianity of each arc contribution. 
For the Caldeira--Leggett model, we will calculate the covariance matrix and compare our result with the constant-switching case.

\section{Example: Damped Jaynes--Cummings model}\label{sec:JC model}

Let us apply our main result \eqref{eq:main solution} to a concrete model. 
We first consider the damped Jaynes--Cummings model, which consists of a two-level system (a ``qubit'') coupled to a single-mode bosonic field. 
The equation of motion for the qubit's quantum state is a first-order homogeneous integro-differential equation, and it is known to be exactly solvable when a constant switching function is employed. 
Hence, it is a good starting point for testing our main result \eqref{eq:main solution}. 
In this section, we compare our result to the exact solution and show that our solution using the train-of-delta method converges to the exact solution in the continuum limit (i.e., the number of delta switchings $N$ increases and the interval $T/N$ between adjacent deltas gets smaller). 
We also demonstrate that the diagrams consisting of single arcs connecting adjacent times correspond to the Markovian dynamics.

\subsection{Solving the master equation}\label{subsec:JC solving QME}

The damped Jaynes--Cummings model describes the coupling between a qubit (a two-level system consisting of a ground $\ket{g}$ and excited state $\ket{e}$) and a single cavity mode. 
We denote the vacuum and one-particle state by $\ket{0}$ and $\ket{1}$, respectively. 
We assume that the transition is restricted to $\ket{g,0}\to \ket{g,0}$ and $\ket{e,0}\leftrightarrow \ket{g,1}$, namely, at most a single excitation is allowed. 
Starting with the initial state $\alpha_0 \ket{g,0} + \alpha_1(0) \ket{e,0}$, the Schr\"odinger equation in the interaction picture yields a first-order homogeneous integro-differential equation \cite{breuer2002theory}: 
\begin{align}
    \dfrac{\dd \alpha_1(t)}{\dd t}
    +
    \int_0^t \dd t'\,
    \Sigma(t,t') \alpha_1(t')=0\,, \label{eq:Schrodinger eq JC model}
\end{align}
where $\Sigma(t,t')\equiv \chi(t) \chi(t') \Gamma(t-t')$ is the memory kernel with 
\begin{align}
    \Gamma(t-t')
    =
        \dfrac{\kappa \Lambda}{2} e^{-\Lambda (t-t')}\,. \label{eq:JC Gamma 1}
\end{align}
Here, $\kappa$ is the coupling constant and $\Lambda$ is the characteristic spectral width of the environment in the frequency domain.

Since $\Gamma(t-t')$ is already stationary, the memory kernel $\Sigma(t,t')$ becomes time-translation invariant if we choose a constant switching function $\chi(t)=1$. 
In this case, the exact solution to the integro-differential equation \eqref{eq:Schrodinger eq JC model} with \eqref{eq:JC Gamma 1} is \cite{breuer2002theory}
\begin{align}
    \alpha_1(t)
    &=
        \alpha_1(0) 
        e^{-\Lambda t/2}
        \kagikako{
            \cosh \dfrac{Dt}{2}
            +
            \dfrac{\Lambda}{D}
            \sinh \dfrac{Dt}{2}
        }\,, \label{eq:JC exact sol}
\end{align}
where $D\equiv \sqrt{\Lambda^2 - 2\kappa \Lambda}$.

For a time-dependent coupling, it is challenging to obtain an analytic solution as the memory kernel is nonstationary. 
We thus apply our solution \eqref{eq:main solution} derived by the delta-coupling method to the Schr\"odinger equation \eqref{eq:Schrodinger eq JC model}. 
We have $\Xi_t=\Xi_i=0$ since it is a homogeneous equation, and $\mathcal O^{(0)}(t)=\alpha_1(0)$. 
Furthermore, the free Green function in this case is $G_0(t)=1$. 
Therefore, we obtain 
\begin{align}
    \alpha_1(t)
    &=
        \mathcal T(t)
        \alpha_1(0)\,,
    \quad
    \mathcal T(t)
    \equiv 
        \mathcal T(t,0)
    \coloneqq
        1 
        + 
        \sum_{l,i=1}^N K_{t,t_l} \mathsf K_{li}\,, \label{eq:JC pure delta sol}
\end{align}
where 
\begin{align}
    K_{t,t_l}
    &=
        -\dfrac{T^2}{N^2}
        \sum_{k=1}^N 
        \Theta(t-t_k) \Sigma(t_k, t_l)
        \Theta(t_k -t_l)\,. 
        \label{Eq: JC polynomial}
\end{align}

\begin{figure}[t]
\centering
\includegraphics[width=\linewidth]{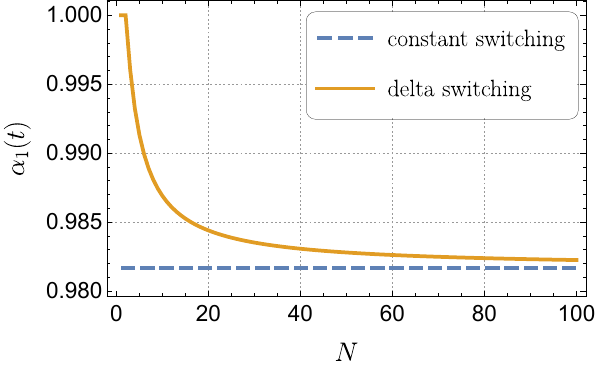}
\caption{
Comparison between the solutions to the integro-differential equation \eqref{eq:Schrodinger eq JC model} with a constant switching (dashed line) and our train of delta switchings (solid curve).
The horizontal axis is taken to be the number of delta switchings $N$.
Here, we chose $\kappa/\Lambda=0.1$.
The function \eqref{eq:JC pure delta sol} at fixed time $\Lambda T=1$ asymptotes to \eqref{eq:JC exact sol} at $\Lambda t=1$ as the number of delta switchings $N$ increases. 
}
\label{fig:JC alpha}
\end{figure}

In Fig.~\ref{fig:JC alpha}, we demonstrate that our solution \eqref{eq:JC pure delta sol} with $\chi_k=1$ ($\forall k$) well approximates the exact solution \eqref{eq:JC exact sol} in the continuum limit $N\to \infty$. 
We see that our solution (solid line) at $t/\Lambda=1$ converges to the exact solution (dashed line) as the number of deltas $N$ increases. 

We note that our Dirac-delta switching method converges to the exact solution for $N \rightarrow \infty$, as long as the formal solution given by the Dyson series converges~\cite{Jose.train.delta.2024}. 
Extra care is needed in numerical calculations in which $N$ is finite.

\subsection{Non-markovianity in the Jaynes--Cummings model}\label{subsec:nonmarkov JC model}

In this subsection, we show that diagrams (Fig.~\ref{fig:N2diagram}) built solely from arcs connecting adjacent times encode Markovian dynamics, whereas long-lived arcs exhibit non-Markovian behavior.

Let $\rho\ts{s}(t)$ be the qubit's quantum state at time $t$. 
From Eq.~\eqref{eq:JC pure delta sol}, the explicit form of the quantum channel $\mathcal E_{t,0}: \rho\ts{s}(0) \mapsto \rho\ts{s}(t)$ and its Kraus representation is given as follows:
\begin{align}
    \mathcal{E}_{t,0} \left[\rho\ts{s}(0)\right] 
    &= 
        \sum_{i=1}^{2} 
        E_i (t) \rho\ts{s} (0) E_i^\dag(t) ,\label{Eq: quantum channel for JC model}\\
    \rho\ts{s}(0) 
    &= 
        \begin{bmatrix}
            |\alpha_1 (0)|^2 & \alpha_0^* \alpha_1 (0) \\
            \alpha_0 \alpha_1^* (0) & 1-|\alpha_1 (0)|^2 
        \end{bmatrix},
\end{align}
with the Kraus operators 
\begin{align}
    E_1 (t)
    \coloneqq
    \begin{bmatrix}
        \mathcal T (t) &0  \\
         0& 1
    \end{bmatrix},
    \quad 
    E_2 (t) 
    \coloneqq 
    \begin{bmatrix}
        0 &0 \\
         \sqrt{1-|\mathcal T (t)|^2} & 0
    \end{bmatrix},
\end{align}
which satisfies the completeness relation 
\begin{align}
    \sum_{i=1}^2 E_i^\dagger (t) E_i(t) = \id\,.
\end{align}

The dynamics $\mathcal E_{t,0}$ is Markovian if it satisfies the so-called semi-group property of the quantum channel, which is a condition $(\mathcal{E}_{t,t'} \circ \mathcal{E}_{t',0}) [\rho\ts{s}(0)] = \mathcal{E}_{t,0} [\rho\ts{s}(0)],$ where $t>t'\geq 0$ (see Appendix~\ref{app:Markovian review} for a brief review). 
The semi-group property for $\alpha_1(t)=\mathcal T(t) \alpha_1(0)$ reduces to the relation $\mathcal T (t) = \mathcal T (t,t') \mathcal T (t')$. 
For the Jaynes--Cummings model, this relation holds when the memory propagation, pictorially represented in Fig.~\ref{fig:N2diagram}, is restricted to nearest-neighbor propagation $t_l \xrightarrow{\Sigma} t_{l+1}$ (see Appendix~\ref{sec:Markov JC} for a proof). 
Furthermore, in the next section we show that the Caldeira--Leggett model exhibits the same behavior under this restriction.

\begin{figure*}[t]
\centering
\includegraphics[width=\linewidth]{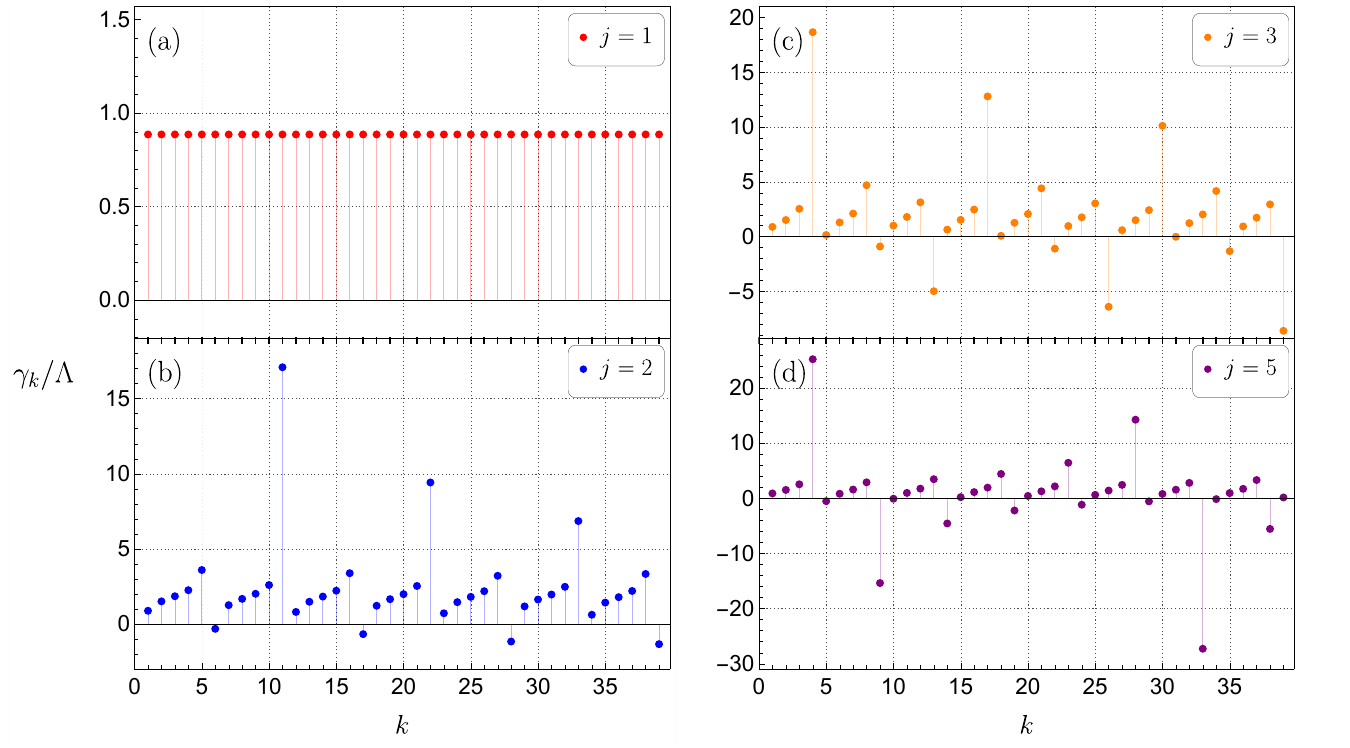}
\caption{
A measure of non-Markovianty of quantum processes.
Non-Markovian behavior arises during the time intervals with $\gamma_{k}/\Lambda<0$, while the dynamics is Markovian otherwise.
Here, we chose $\kappa/\Lambda=5/2$, $N=40$ and $\Lambda T=30$.
The integer $j$ denotes the extent of memory propagation $t_l \xrightarrow{\Sigma} t_{l+j}$ allowed in the dynamics. 
$j=1$ in (a) allows only nearest-neighbor propagation, whereas $j\geq 2$ in (b)-(d) permits propagation up to $j$ steps away.
Note that the time evolution is Markovian for $t\geq t_{40}$, where $k=40$ denotes the time at which the interaction ends. Therefore, we exclude the point $k=40$ from our analysis.
}
\label{fig:JCFig}
\end{figure*}

To numerically demonstrate that the Markovianity of the dynamics arises from nearest-neighbor propagation, let us introduce a measure for non-Markovianity. 
We employ the measure proposed by Rivas, Huelga, and Plenio (RHP) in \cite{Rivas.nonMarkovMeasure.2010}, which is based on the fact that non-Markovian dynamics is completely-positive (CP) indivisible \cite{Breuer.nonMarkovMeasure.2009, Rivas.nonMarkovMeasure.2010, Vega.nonMarkov.2017}.
In the case of the damped Jaynes--Cummings model presented above, the RHP measure $\cal I$ becomes \cite{Rivas.nonMarkovMeasure.2010, Zeng.EquivalofnonMarkovMeasure.2011}
\begin{align}
    \mathcal I
    &\coloneqq
        - \int_{\gamma(t)<0}
        \gamma(t) \dd t\,, \label{eq:RHP measure}
\end{align}
where $\gamma(t)$ is the decay rate\footnote{Note that there are other measures for non-Markovianity, such as the $g$ function and the trace measure. We comment on their relation to the decay rate in the case of finite $T/N$ in Appendix~\ref{app:nonMarkov measure}.
}  given by \cite{breuer2002theory}
\begin{align}
    \gamma(t)
    &\equiv 
        -2 \text{Re}
        \kagikako{
            \dfrac{\dot \alpha_1(t)}{\alpha_1(t)}
        }\,.
        \label{Eq: time-dependent decay rate}
\end{align}
If $\mathcal I=0$ (i.e., $\gamma(t)\geq0$) then the dynamics is Markovian, while it is non-Markovian for $\mathcal I>0$ (i.e., $\gamma(t)<0$). 
The RHP measure for our delta method reads 
\begin{align}
    \mathcal I
    &=
        - \dfrac{T}{N} \sum_{k=1}^N 
        \min 
        \left\{ 
            0,\, 
            \gamma_k
        \right\}\,, \notag \\
    \gamma_k 
    &\equiv
        -2
        \dfrac{T}{N}
        \text{Re}
        \kagikako{
            \dfrac{ \sum_{j=1}^{k} \Sigma(t_{k+1}, t_j) \alpha_1(t_j) }{ \kako{ 1+ \sum_{l, i=1}^N K_{kl} \mathsf K_{li} } \alpha_1(0) }
        }. \label{eq:gamma_k}
\end{align}
Here, $\gamma_k$ represents the amplitude of $\gamma(t)$ at time $t=t_k$ when a delta-switching is employed. 
The formula above suggests that it is necessary to have $\gamma_k<0$ for $\mathcal E_{t,0}$ to be non-Markov. 
We note that the RHP measure takes this simple form in \eqref{eq:gamma_k} only in the regime $\kappa T/N\ll 1$. 
Otherwise, the expression for $\cal I$ is significantly more complicated. 
See Appendix~\ref{app:nonMarkov measure} for details.

Figure~\ref{fig:JCFig} shows the decay rate $\gamma_k$ at time $t_k$ given in Eq.~\eqref{eq:gamma_k} when the memory propagation $t_l \xrightarrow{\Sigma} t_{l+j}$ is allowed up to $j$ steps away for each $l=1,2,\ldots,k$. 
For example, $j=3$ in Fig.~\ref{fig:JCFig}(c) depicts $\gamma_k$ when $t_l \xrightarrow{\Sigma} t_{l+1}$, $t_l \xrightarrow{\Sigma} t_{l+2}$, and $t_l \xrightarrow{\Sigma} t_{l+3}$ are allowed. 
Hence, Fig.~\ref{fig:JCFig}(a) with $j=1$ permits only nearest-neighbor propagation. 
As discussed in Sec.~\ref{subsec:pictorial}, diagrams composed only of arcs connecting adjacent times [Fig.~\ref{fig:JCFig}(a)] correspond to Markovian dynamics, as $\gamma_k$ is always non-negative. 
On the other hand, the presence of long-lived arcs [Figs.~\ref{fig:JCFig}(b)-(d)] shows non-Markovian behavior when $\gamma_k<0$.


\section{Example: Damped harmonic oscillator}\label{sec:QLE}

In this section, we consider another example: a damped harmonic oscillator. 
This consists of a harmonic oscillator (as a principal system) coupled to a collection of harmonic oscillators (an environment). 
Instead of working in the Schr\"odinger picture, we consider the observables (the quadratures $Q$ and $P$ of the system) in the Heisenberg picture. 
The Heisenberg equation of motion for $Q(t)$ is known as the quantum Langevin equation (QLE), which is a second-order inhomogeneous integro-differential equation. 
We will solve the QLE with a nonstationary memory kernel\footnote{There have been few studies on the QLE with time-dependent coupling (i.e., non-stationary QLE). 
For example, Refs.~\cite{Freitas.Fundamental.2017, Freitas.Cooling.2018, Carrega.Engineering.2022, Cavaliere.Dynamical.2022} apply Floquet theory to study the asymptotic behavior of periodically driven systems obeying the QLE. 
Moreover, in the field of coarse-grained molecular dynamics simulation, the generalized Langevin equation (GLE) in the Mori-Zwanzig theory has been extensively studied. In recent years, non-stationary GLE has been analyzed by using the so-called memory reconstruction method \cite{Meyer.GLE.2017, Meyer.NonMarkovian.2020, Klippenstein.Molecular.review.2021}.} by applying our method \eqref{eq:main solution}, and demonstrate how to deal with the two-time correlation functions $\braket{Q(t)Q(t')}$ as well as the covariance matrix, which is directly related to the quantum state of the system.

\subsection{Solving the QLE}\label{subsec:Solving QLE}
We begin by considering the Caldeira--Leggett model, where the environment is modeled as a collection of $\mathcal{N}$ independent quantum harmonic oscillators. 
Note that, at this stage, the environment contains a finite number of oscillators. 
After solving the QLE, we calculate the correlation functions and the covariance matrix of the system and then take the limit $\mathcal{N}\to \infty$.

Consider a system modeled by a quantum harmonic oscillator of frequency $\Omega$ (and mass $M=1$), whose quadratures are denoted by $Q$ and $P$. 
The free Hamiltonian $H\ts{s,0}$ is given by 
\begin{align}
    H\ts{s,0}
    &=
        \dfrac{P^2}{2} 
        + 
        \dfrac{\Omega^2 }{2} Q^2\,.
\end{align}
The environment is composed of $\mathcal{N}$ quantum harmonic oscillators, each labeled by $j\in \{ 1,2, \ldots, \mathcal{N} \}$. 
For each unit mass oscillator, the frequency and quadratures are given by $\omega_j$, $q_j$, and $p_j$, respectively, and the free Hamiltonian $H\ts{E,0}$ is 
\begin{align}
    H\ts{E,0}
    &=
        \sum_{j=1}^\mathcal{N}
        \kako{
            \dfrac{p_j^2}{2}
            +
            \dfrac{\omega_j^2}{2} q_j^2
        }. 
\end{align}
Assuming that the system couples identically to each oscillator in the environment, the interaction between the system and the environment is described by the interaction Hamiltonian 
\begin{align}
    H\ts{int}(t)
    &=
        -c(t) Q
        \otimes \sum_{j=1}^{\mathcal{N}} 
        q_j\,,
\end{align}
where $c(t)$ represents the time-dependent coupling between the system and the environment.

To derive the QLE, consider the Heisenberg equations of motion for the quadratures in the Heisenberg picture. 
These lead to coupled second-order differential equations: 
\begin{subequations}
\begin{align}
    &\ddot{Q}(t) + \Omega^2 Q(t) 
    =
        c(t)
        \sum_{j=1}^{\mathcal N} q_j(t)\,, \label{eq:Heisenberg EOM system}\\
    &\ddot{q}_j(t) 
    + 
    \omega_j^2 q_j(t)
    =
        c(t) 
        Q(t)\,.\label{eq:Heisenberg EOM env}
\end{align}
\end{subequations}
The equation of motion for oscillator-$j$ in \eqref{eq:Heisenberg EOM env} can be solved as 
\begin{align}
    q_j(t)
    &= 
        q_j^{(\text h)}(t) \notag \\
        &
        + 
        \int_0^\infty \dd t'\,
        \dfrac{ \sin(\omega_j (t-t'))  }{\omega_j} 
        \Theta(t-t')
        c(t') Q(t')\,,
\end{align}
where $q_j^{(\text h)}(t)$ is the homogeneous solution to the equation. 
Inserting this solution into the right-hand side of \eqref{eq:Heisenberg EOM system} gives the QLE: 
\begin{align}
    \ddot{Q}(t)
    + \Omega^2 Q(t)
    + \int_0^t \dd t' \,
    \Sigma(t,t') Q(t')
    &=
        \zeta(t)\,, \label{eq:nonstationary QLE} 
\end{align}
where the memory kernel $\Sigma(t)$ is defined by\footnote{For the QLE, we will adopt the convention that a minus appears in front of $\Gamma(t)$.} 
\begin{subequations}
\begin{align}
    &\Sigma(t,t')
    \coloneqq
        - \chi(t) \chi(t')
        \Gamma(t-t') \,, \\
    &\Gamma(t)
    \coloneqq
        \sum_{j=1}^{\mathcal{N}}
        \dfrac{c^2}{ \omega_j}
        \sin(\omega_j t)\,, \label{eq:Gamma discrete}
\end{align}
\end{subequations}
and the noise term is given by 
\begin{align}
    \zeta(t)
    &\coloneqq
        \chi(t) \xi(t)\,,
    \quad
    \xi(t)
    \equiv 
        c 
        \sum_{j=1}^{\mathcal{N}}
        q_j^{(\text{h})}(t)\,. \label{eq:noise term}
\end{align}

As explained in Sec.~\ref{subsec:solving integro-diff eq}, the integro-differential equation can be solved when the memory kernel is stationary. 
In this case, $\Gamma(t,t')$ is already stationary, so $\Sigma(t,t')$ becomes time-translation invariant if the system constantly couples to the environment: $\chi(t)=1$. 

In this scenario, the resulting algebraic equation in the Laplace domain leads to 
\begin{align}
    \tilde Q(z)
    &=
        z \tilde G(z) Q(0)
        + 
        \tilde G(z) \dot Q(0)
        +
        \tilde G(z) \tilde \zeta(z)\,, \notag 
\end{align}
where $\tilde G(z) \coloneqq ( z^2 + \Omega^2 + \tilde \Sigma(z) )^{-1}$. 
Notice that $\zeta(t)=\xi(t)$. 
The inverse Laplace transformation gives us the well-known general solution 
\begin{align}
    Q(t)
    &=
        \dot G(t) Q(0)
        + 
        G(t) \dot Q(0)
        + 
        \int_0^t \dd t'\,
        G(t-t') \zeta(t')\,.\label{eq:cont sol}
\end{align}
In practice, to compute quantities such as the correlation functions $\braket{Q(t) Q(t')}$, one needs the explicit form of $G(t)$, the inverse Laplace transform of $\tilde G(z)$. 
This requires further assumptions, for example, specifying an explicit form of the spectral density (see, e.g., \cite{Boyanovsky.Langevin.2017}), which we introduce later.

We now consider a general switching function $\chi(t)$ and apply our result \eqref{eq:main solution}. 
The free solution $Q^{(0)}_t$ and the free Green function $G_0(t)$ for a quantum harmonic oscillator are given by 
\begin{align}
    &Q^{(0)}_t
    =
        Q(0) \cos (\Omega t)
        + 
        \dfrac{P(0)}{ \Omega} 
        \sin (\Omega t)\,, \notag \\
    &G_0(t)
    =
        \dfrac{\sin(\Omega t)}{\Omega}\,,\notag 
\end{align}
and thus Eq.~\eqref{eq:main solution} with these functions is the exact solution to the QLE~\eqref{eq:nonstationary QLE} when a compactly supported switching function is described by the train of Dirac deltas~\eqref{eq:train of delta}. 
For the sake of later use, we write the solution to the QLE in the following form: 

\begin{widetext}
\begin{align}
    Q(t)
    &=
        \kagikako{
            \cos (\Omega t)
            +
            \sum_{l,i=1}^N K_{t, t_l} 
            \mathsf{K}_{li} \cos (\Omega t_i)
        }
        Q(0) 
        + 
        \kagikako{
            \dfrac{\sin(\Omega t)}{ \Omega  }
            +
            \sum_{l,i=1}^N K_{t, t_l} 
            \mathsf{K}_{li} \dfrac{\sin(\Omega t_i)}{ \Omega  } 
        }
        P(0)  \notag \\
        &\quad +
        \sum_{k=1}^N 
        \kagikako{
            \dfrac{\sin[\Omega (t-t_k)]}{ \Omega  }
            \Theta(t-t_k)
            +
            \sum_{l,i=1}^N K_{t, t_l} 
            \mathsf{K}_{li} \dfrac{\sin[\Omega (t_i-t_k)]}{ \Omega  } 
            \Theta(t_i-t_k)
        }
        \zeta(t_k)\,,
    \notag \\
    &\equiv
        G[f_Q](t) Q(0)
        +
        G[f_P](t) P(0)
        +
        \sum_{k=1}^N 
        G[f_\zeta^{(k)}](t) \zeta(t_k)\,,\quad \forall t>0\,,\label{eq:main result} 
\end{align}
\end{widetext}
where $G[f]$ is a functional defined as 
\begin{subequations}
\begin{align}
    &G[f](t)
    \coloneqq
        f(t) + \sum_{l,i=1}^N K_{t, t_l} 
        \mathsf{K}_{li} f(t_i) \,,\\
    &f_Q(t) 
    \coloneqq
        \cos (\Omega t)\,, \\
    &f_P(t)
    \coloneqq
        \dfrac{\sin(\Omega t)}{ \Omega  }\,, \\
    &f_{\zeta}^{(k)}(t)
    \coloneqq
        \dfrac{\sin[\Omega (t-t_k)]}{\Omega }
        \Theta(t-t_k)\,.
\end{align}\label{eq:main result functions}
\end{subequations}
In what follows, we denote $G_{f}(t) \equiv G[f](t)$ for brevity.

Equation~\eqref{eq:main result} is the exact solution to the QLE~\eqref{eq:nonstationary QLE} when a compactly supported switching function is described by the train of Dirac deltas~\eqref{eq:train of delta}. 
Note that this solution is also valid for intermediate times $t\in (0,T)$, as the Heaviside step functions naturally eliminate irrelevant terms. 
Moreover, at each discrete time $t_l$ for $l\in \{ 1,2,\ldots, N \}$, the solution $Q(t=t_l)$ reduces to Eq.~\eqref{eq:main result special case}, which can be expressed using a single function $\mathcal G(t)$ as 
\begin{align}
    Q(t_l)
    &=
        \dot{\mathcal{G}}(t_l) Q(0) 
        + 
        \mathcal{G}(t_l) P(0)  \notag \\
        &\quad 
        +
        \sum_{k=1}^N 
        \mathcal{G}(t_l-t_k)\Theta(t_l-t_k) \zeta(t_k)\,, \label{eq:main result special case v2}
\end{align}
where 
\begin{align}
    \mathcal{G}(t)
    &\coloneqq
        \dfrac{\sin(\Omega t)}{\Omega}
        +
        \sum_{l,i=1}^N K_{t, t_l} 
        \mathsf{K}_{li} \dfrac{\sin(\Omega t_i)}{\Omega}\,.
\end{align}
This is consistent with the well-known solution for constant switching, as the continuum limit of Eq.~\eqref{eq:main result special case v2} reduces to Eq.~\eqref{eq:cont sol}. 

\subsection{Two-time correlation functions and the covariance matrix}\label{subsec:correlation function and CM}

In the literature, two-time correlation functions such as $\braket{Q(t) Q(t')}$ and $\braket{P(t) P(t')}$, as well as their asymptotic values are the primary focus. 
A related quantity of interest, especially in Gaussian quantum mechanics \cite{serafini.QCV}, is the \textit{covariance matrix} $\mathbb{V}$, which determines a quantum state. 
In the following, we study these by using our solution to the QLE \eqref{eq:main result}.

\subsubsection{Two-time correlation functions}

In the Heisenberg picture, consider the system's quadrature $\bm R(t)\equiv [Q(t), P(t)]^\intercal$. 
The two-time correlation function is defined by 
\begin{align}
    \braket{R_i(t) R_j(t')}
    &\equiv 
        \braket{R_i(t) R_j(t')}_{\rho\ts{tot}(0)} \notag \\
    &\coloneqq
        \Tr[ \rho\ts{tot}(0) R_i(t) R_j(t') ]
    \quad (i,j \in \{ 1,2 \})\,.
\end{align}
We note that the expectation values for $\bm R(t)$ are taken with respect to the total initial state $\rho\ts{tot}(0)$, as $Q(t)$ and $P(t)$ are considered to be observables on the total Hilbert space [see Eq.~\eqref{eq:main result}]. 
On the other hand, the expectation values for the initial quadratures $Q(0)$ and $P(0)$, and those for the environment's observable $\zeta(t)\equiv \chi(t) \xi(t)$ should be understood as $\braket{R_j(0)}\equiv \braket{R_j(0)}_{\rho\ts{s}(0)}$ and $\braket{\zeta(t)}\equiv \braket{\zeta(t)}_{\rho\ts{E}(0)}$, where $\rho\ts{s}(0)$ and $\rho\ts{E}(0)$ are the initial states of the system and the environment, respectively. 
In what follows, we omit the subscripts $\rho\ts{tot}(0)$, $\rho\ts{s}(0)$, and $\rho\ts{E}(0)$.

Let us focus on $\braket{Q(t)Q(t')}$ and explicitly write in terms of the functions in \eqref{eq:main result functions}. 
We assume that the initial joint state is a product state, 
\begin{align}
    \rho\ts{tot}(0)
    =
        \rho\ts{s}(0) \otimes \rho\ts{E}(0)\,, 
\end{align}
and that the environment's one-point correlation function is zero, $\braket{\zeta(t)}=0$, and any $n$-point correlation functions, $\braket{\zeta(t_1)\ldots \zeta(t_n)}$, can be written as products of the environment's two-time correlation functions, $\braket{\zeta(t) \zeta(t')}$. 
The thermal Gibbs state is one of the examples. 
From \eqref{eq:main result}, the two-time correlation function $\braket{Q(t)Q(t')}$ reads 
\begin{align}
    \braket{Q(t) Q(t')}
    &=
        G_{f_Q}(t) G_{f_Q}(t') 
        \braket{ Q^2(0) } \notag \\
        &\quad
        +
        G_{f_P}(t) G_{f_P}(t') 
        \braket{ P^2(0) } \notag\\
        &\quad
        +
        G_{f_Q}(t) G_{f_P}(t') \braket{ Q(0) P(0) } \notag \\
        &\quad
        +
        G_{f_P}(t) G_{f_Q}(t') \braket{ P(0) Q(0) } \notag \\
        &\quad
        +
        \sum_{k,k'=1}^N 
        G_{f_\zeta^{(k)}}(t) G_{f_\zeta^{(k')}}(t')
        \braket{\zeta(t_k) \zeta(t_{k'})}\,. \label{eq:QQ corr}
\end{align}

As we point out later, the delta switchings allow us to write the correlation functions in terms of discretized data, $\Gamma(t_k)$ and $\braket{\zeta(t_k)\zeta(t_{k'})}$. 
This suggests that the discrete-time Fourier transform is preferred (as opposed to the continuous-time Fourier transform) for defining quantities in the frequency domain such as the spectral density.

\subsubsection{The covariance matrix}
In continuous-variable quantum mechanics, a quantum state is called \textit{Gaussian} if it is determined solely by the first and second moments. 
We denote the first moment by $\bar{\bm R}\equiv \braket{\bm R}$. 
For the second moments, one typically considers the covariance matrix $\mathbb{V}(t)$---a real symmetric, positive-definite matrix---for describing the second moments, and it is given by \cite{Adesso.CV.review.2014}
\begin{align}
    \mathbb{V}_{ij}(t)
    &\coloneqq
        \braket{ \{ R_i(t), R_j(t) \} }
        - 
        2 \braket{R_i(t)} \braket{R_j(t)}\,.
\end{align}
The vacuum state, the thermal Gibbs state, and the squeezed state are examples of the Gaussian state with vanishing first moments, and the coherent state is a Gaussian state with a nonvanishing first moment. 
Once these statistical moments are known, we can obtain the corresponding Gaussian state $\rho\ts{G} \equiv \rho\ts{G}[\bar{\bm R}, \mathbb V]$.

A particularly useful tool for studying the covariance matrix in Gaussian systems is the \textit{Gaussian operations}. 
These are the completely-positive (CP) maps that transform Gaussian states to Gaussian states \cite{serafini.QCV}. 
To illustrate this, suppose a joint system is initially prepared in a product of Gaussian states, $\rho\ts{tot}(0)=\rho\ts{s}(0) \otimes \rho\ts{E}(0)$, which is also Gaussian. 
A unitary time-evolution operator generated by a Hamiltonian composed of a linear and a quadratic term in $R_i$ is an example of the Gaussian operation known as the Gaussian unitary transformation \cite{serafini.QCV}. 
The Hamiltonian in the Caldeira--Leggett model used in this paper indeed generates such a unitary transform, and this fact does not change when a train of delta-switchings is employed. 
Moreover, a partial trace of a Gaussian joint state is also a Gaussian operation. 
Thus, the system's reduced quantum state $\rho\ts{s}(t)=\Tr\ts{E}[ U(t) \rho\ts{tot}(0) U^\dag(t) ]$ after the time-evolution remains to be Gaussian. 
This fact can be utilized to characterize the evolution of the statistical moments. 
Assuming the initial joint state is separable Gaussian and they evolve under a trace-preserving Gaussian operation (hence, it is a CPTP map), the first moment and the covariance matrix evolve as \cite{serafini.QCV}
\begin{subequations}
\begin{align}
    &\bar{\bm R}
    \mapsto 
        \mathbb{T} \bar{\bm R}\,, \\
    &\mathbb{V}
    \mapsto 
        \mathbb{T} \mathbb{V} \mathbb{T}^\intercal 
        + 
        \mathbb{N}\,,
\end{align} \label{eq:R and V under UDM}
\end{subequations}
where $\mathbb{T}$ and $\mathbb{N}$ are real square matrices satisfying the condition that emerges from the uncertainty principle, 
\begin{align}
    \mathbb{N} 
    + \ii \bm \Omega
    \geq \ii \mathbb{T} \bm \Omega \mathbb T^\intercal\,,
\end{align}
where $\bm \Omega$ is the symplectic form that appears in the canonical commutation relations: $[R_i, R_j]=\ii \Omega_{ij}$.

One can prove that the Markovianity of a quantum channel is encoded in the properties of $\mathbb T$ and $\mathbb N$. 
Let $\mathcal{E}_{t,s}: \rho(s) \mapsto \rho(t)$ be a Gaussian CPTP map. 
If $\mathcal{E}_{t,s}$ is a Markovian dynamical map, then the associated matrices $\mathbb T$ and $\mathbb N$ obey: 
\begin{align}
    \mathbb T_{t+s}
    &=
        \mathbb T_t \mathbb T_s\,,
    \quad
    \mathbb N_{t+s}
    =
        \mathbb T_t \mathbb N_s \mathbb T_t^\intercal 
        + 
        \mathbb N_t\,, \label{eq:Markovianity conditions for T N}
\end{align}
where $\mathbb{T}_t$ and $\mathbb N_t$ are the matrices associated with $\mathcal{E}_{t,0}$. 
We prove these properties in Appendix~\ref{app:Markov}.

The evolution for $\bar{\bm R}$ and $\mathbb{V}$ given in Eq.~\eqref{eq:R and V under UDM} holds for generic CPTP Gaussian operations, as long as the initial joint state is separable Gaussian. 
We now apply this to our delta-switching case when the system is initially prepared in a Gaussian state. 
In particular, we assume that the environment's initial Gaussian state has a zero first moment, $\braket{\zeta(0)}=0$.

From the result in \eqref{eq:QQ corr} and other correlation functions such as $\braket{ \{ Q(t), P(t) \} }$ allow us to obtain the matrices $\mathbb T_t$ and $\mathbb N_t$ for $t>T$ as 
\begin{align}
    &\mathbb V(t)
    =
        \mathbb T_t \mathbb V(0) \mathbb T^\intercal_t + \mathbb N_t\,, \notag \\
    &\mathbb T_t
    =
        \begin{bmatrix}
            G_{f_Q}(t) & G_{f_P}(t) \\
            \dot G_{f_Q}(t) & \dot G_{f_P}(t)
        \end{bmatrix}\,, \\
    &\mathbb N_t
    =
        \begin{bmatrix}
            N_{QQ}(t) & N_{QP}(t) \\
            N_{PQ}(t) & N_{PP}(t)
        \end{bmatrix}, 
\end{align}

where 
\begin{subequations}
\begin{align}
    N_{QQ}(t)
    &=
        \sum_{k,k'=1}^N 
        G_{f_\zeta^{(k)}}(t) G_{f_\zeta^{(k')}}(t)
        \nu_{k,k'}\,, \\
    N_{QP}(t)
    &=
        \sum_{k,k'=1}^N 
        G_{f_\zeta^{(k)}}(t) \dot G_{f_\zeta^{(k')}}(t)
        \nu_{k,k'}\,, \\
    N_{PQ}(t)
    &=
        \sum_{k,k'=1}^N 
        \dot G_{f_\zeta^{(k)}}(t) G_{f_\zeta^{(k')}}(t)
        \nu_{k,k'}\,, \\
    N_{PP}(t)
    &=
        \sum_{k,k'=1}^N 
        \dot G_{f_\zeta^{(k)}}(t) \dot G_{f_\zeta^{(k')}}(t)
        \nu_{k,k'}\,,
\end{align}\label{eq:Noises for delta}
\end{subequations}
and $\nu_{k,k'}\equiv\braket{\{\zeta(t_k) ,\zeta(t_{k'})\}}$.
It turns out that the matrices $\mathbb T_t$ and $\mathbb N_t$ for the case of delta-switchings do not satisfy the Markovianity conditions \eqref{eq:Markovianity conditions for T N} in general. 
However, one can reduce the dynamics to Markovian by applying the Born--Markov approximation. 
We demonstrate this fact in Appendix~\ref{app:BornMarkov} for an environment with the so-called Lorentz-Drude spectral density described in the next section.

\subsection{Demonstration with the Lorentz--Drude spectral density}\label{subsec:spectral density}

Our solution to the nonstationary QLE in \eqref{eq:main result} was derived under the assumption that the environment consists of a finite collection of quantum harmonic oscillators. 
Nevertheless, we can extend this result to the continuum limit by introducing the \textit{spectral density} $\sigma(\omega)$ of the environment. 
In the traditional QLE with a continuous constant coupling, the spectral density is defined via the Fourier transform of the memory kernel. 
In contrast, we illustrate that the spectral density should be defined by using the discrete-time Fourier transform (DTFT) when a train of delta switchings is employed.

\subsubsection{Introducing the spectral density}\label{subsubsec:conventional spectral density}

Let us introduce the traditional spectral density used when a continuous constant coupling is considered. 
Our first aim is to rewrite the quantities that appear in $\braket{R_i(t)}$ and $\braket{R_i(t) R_j(t')}$ in terms of the spectral density $\sigma(\omega)$. 
Consider the time-translation invariant part $\Gamma(t)$ of the memory kernel. 
For a finite collection of quantum harmonic oscillators, $\Gamma(t)$ is given by \eqref{eq:Gamma discrete}. 
We then introduce the spectral density $\sigma(\omega)$ by Fourier transforming $\Gamma(t)$ as\footnote{In this paper, the Fourier transform is denoted by $\hat \Gamma(\omega) \equiv \mathcal{F}[\Gamma(t)]$. This notation should not be confused with that used for linear operators in quantum mechanics.} 
\begin{align}
    \hat \Gamma(\omega)
    &\coloneqq
        \int_{\R} \dd t\,
        \Gamma(t) e^{\ii \omega t}
    \equiv 2 \ii \sigma(\omega)\,, \label{eq:spectral density continuous-time FT}
\end{align}
where 
\begin{align}
    \sigma(\omega)
    &=
        \sum_{j=1}^{\mathcal N}
        \dfrac{\pi}{2}
        \dfrac{c^2 }{\omega_j} 
        \kagikako{
            \delta(\omega - \omega_j)
            -
            \delta(\omega + \omega_j)
        } \label{eq:spectral density discrete}
\end{align}
is the spectral density of the environment composed of a finite collection of harmonic oscillators. 
Note that it has the property $\sigma(\omega)=-\sigma(-\omega)$. 
Then, $\Gamma(t)$ can be expressed using the spectral density as 
\begin{align}
    \Gamma(t)
    &=
        \dfrac{\ii}{\pi} 
        \int_{\R} \dd \omega\,
        \sigma(\omega) e^{-\ii \omega t}\,,
\end{align}
and thereby the memory kernel $\Sigma(t,t')$ reads  
\begin{align}
    \Sigma(t,t')
    &=
        -\chi(t) \chi(t') 
        \dfrac{\ii}{\pi} 
        \int_{\R} \dd \omega\,
        \sigma(\omega) e^{-\ii \omega (t - t')}\,.
\end{align}

We also need to express the quantities related to the noise term $\zeta(t)$, defined in \eqref{eq:noise term}, in terms of $\sigma(\omega)$. 
Instead of establishing a direct relationship between $\zeta(t)$ and $\sigma(\omega)$, we express the correlation functions $\braket{ \zeta(t) \zeta(t') }$ in terms of $\sigma(\omega)$, as these correlations are required for evaluating the two-time correlation functions. 
To this end, we assume that the initial joint state is a product state and that the environment is prepared in a thermal Gibbs state: 
\begin{align}
    \rho\ts{tot}(0)
    &=
        \rho\ts{s}(0) \otimes \rho\ts{E}(0)\,,
    \quad
    \rho\ts{E}(0)
    =
        \dfrac{1}{Z} e^{-\beta H\ts{E}}\,,
\end{align}
where $Z\coloneqq \Tr[ e^{-\beta H\ts{E} }]$ is the partition function and $\beta>0$ is the inverse temperature of the environment. 
Then, one can straightforwardly verify that the one-point and two-point correlation functions read 
\begin{subequations}
\begin{align}
    &\braket{ \zeta(t)}
    =0\,, \\
    &\braket{\zeta(t) \zeta(t')}
    =
        \chi(t) \chi(t') \braket{\xi(t) \xi(t')}\,, \\
    &\braket{\xi(t) \xi(t')}
    =
        \dfrac{1}{\pi} 
        \int_\R \dd \omega\,
        \dfrac{\sigma(\omega)}{ e^{\beta \omega} -1 }
        e^{ -\ii \omega(t-t') }\,. \label{eq:xi correlation}
\end{align}
\end{subequations}
Again, the correlation functions for the environment's observable should be understood as $\braket{\zeta(t)}\equiv \braket{\zeta(t)}_{\rho\ts{E}(0)}$.

By introducing the spectral density $\sigma(\omega)$, each term in the two-time correlation function $\braket{R_i(t) R_j(t')}$ can be rewritten in terms of it. 
This approach allows us to examine a variety of environments, including those with an infinite number of quantum harmonic oscillators. 
The idea is as follows. 
So far, our $\sigma(\omega)$ given in \eqref{eq:spectral density discrete} is the explicit form for a finite collection of quantum harmonic oscillators. 
To consider other types of environments (e.g., the case where $\mathcal{N}\to \infty$), we simply replace the spectral density in \eqref{eq:spectral density discrete} with the one of our interest. 
A widely examined spectral density is the \textit{Lorentz--Drude (LD) spectral density}: 
\begin{align}
    \sigma(\omega)
    &=
        \kappa \dfrac{\omega \Lambda^2}{ \omega^2 + \Lambda^2 }\,, \label{eq:LD spectral density}
\end{align}
where $\kappa>0$ is the coupling constant and $\Lambda>0$ is the cutoff frequency that determines the bandwidth of the environment. 
Moreover, $\Lambda^{-1}$ is the characteristic time scale of the change in the environment. 
Given a cutoff $\Lambda$, the low-frequency regime $\omega \ll \Lambda$ of the LD spectral density mimics the Ohmic system as $\sigma(\omega)\approx \kappa \omega$, which reflects the Markovian dynamics. 
Employing the LD spectral density, we have 
\begin{align}
    \Gamma(t)
    &=
        \kappa \Lambda^2 e^{-\Lambda |t|} \sgn(t)\,. \label{eq:Gamma LD}
\end{align}

\subsubsection{Spectral density for discretized data}\label{subsubsec:descrete spectral density}

When considering continuous (typically constant) switching functions $\chi(t)$, one can compute correlation functions $\braket{R_i(t) R_j(t')}$ using the spectral density defined above. 
In this paper, however, the train of delta switchings requires us to adopt the \textit{discrete-time Fourier transform} (DTFT) \cite{oppenheim1999discrete}---the discrete-time variant of the continuous-time Fourier transform---to define the spectral density.

To see this, consider the two-time correlation functions such as $\braket{Q(t) Q(t')}$. 
As we saw in \eqref{eq:QQ corr}, these correlation functions are composed of the terms such as $\braket{\zeta(t_k) \zeta(t_{k'})}$, which is the noise correlation function evaluated at the delta-switched (discrete) times $t_k$ and $t_{k'}$. 
This suggests that one deals with sequences instead of continuous functions to compute the correlation functions. 
Therefore, instead of employing the continuous-time Fourier transform to define the spectral density and the noise correlations, it is suitable to choose the DTFT for our delta-switched scenario.\footnote{Simply inserting $t=t_k$ and $t'=t_{k'}$ in the noise correlator \eqref{eq:xi correlation} defined using the continuous-time Fourier transform leads to divergence at $k=k'$.}

From now on, we redefine the spectral density $\sigma(\omega)$ by employing the DTFT. 
Let $\{ f_n \}$ be an absolutely summable discrete-time sequence. 
The DTFT of $f_k$ [denoted by $\breve f(\overline\omega)$] is defined as 
\begin{align}
    \breve f(\overline\omega)
    &\coloneqq
        \sum_{n=-\infty}^\infty 
        f_n e^{ \ii \overline\omega n }\,, \label{eq:DTFT}
\end{align}
and the inverse DTFT is given by 
\begin{align}
    f_n
    &=
        \int_{-\pi}^\pi
        \dfrac{\dd \overline\omega}{2\pi}\,
        \breve f(\overline\omega) e^{-\ii \overline\omega n}\,. \label{eq:inverse DTFT}
\end{align}
Note that $\overline\omega$ has units of radians, which relates to the frequency $\omega$ in the Fourier transform via $\overline \omega = \omega T/N$. 
Also $\breve f(\overline\omega)$ is periodic, $\breve f(\overline\omega + 2\pi)=\breve f(\overline\omega)$, where the periodicity $2\pi$ is called the sampling frequency. 
Recalling that a periodic function can be expressed as a Fourier series, the definition of the DTFT \eqref{eq:DTFT} is basically the Fourier series in the frequency domain, and $f_n$ corresponds to the Fourier coefficient.

Using the DTFT defined above, let us consider $\Gamma_k \equiv \Gamma(t_k)$. 
Our aim is to replace the spectral density $\sigma(\omega)$ with the periodic spectral density $s(\overline\omega)$ associated with $\Gamma_k$ as shown in Eq.~\eqref{eq:Gamma_k in terms of s}. 
The idea is as follows. 
Suppose the environment of interest has the memory effect described by the LD spectral density \eqref{eq:LD spectral density}, which gives $\Gamma(t)$ in \eqref{eq:Gamma LD}. 
The train of delta-switchings allows us to compute the two-time correlation functions by using discretized data such as $\Gamma_k$ and 
\begin{align}
    \braket{\zeta_k \zeta_{k'}}
    \equiv 
        \braket{\zeta({t_k}) \zeta(t_{k'})}
    =
        \dfrac{T^2}{N^2} 
        \chi(t_k) \chi(t_{k'})
        \braket{\xi_k \xi_{k'}}\,.
\end{align}
Here $\braket{\xi_k \xi_{k'}}\equiv \braket{\xi({t_k}) \xi(t_{k'})}$ and we used Eq.~\eqref{eq:Laplace of zeta}. 
Thus, it is natural to employ the DTFT to express the frequency representations of these discretized data, and we denote the spectral density defined through the DTFT by $s(\overline \omega)$.

\begin{figure*}[t]
\centering
\includegraphics[width=\linewidth]{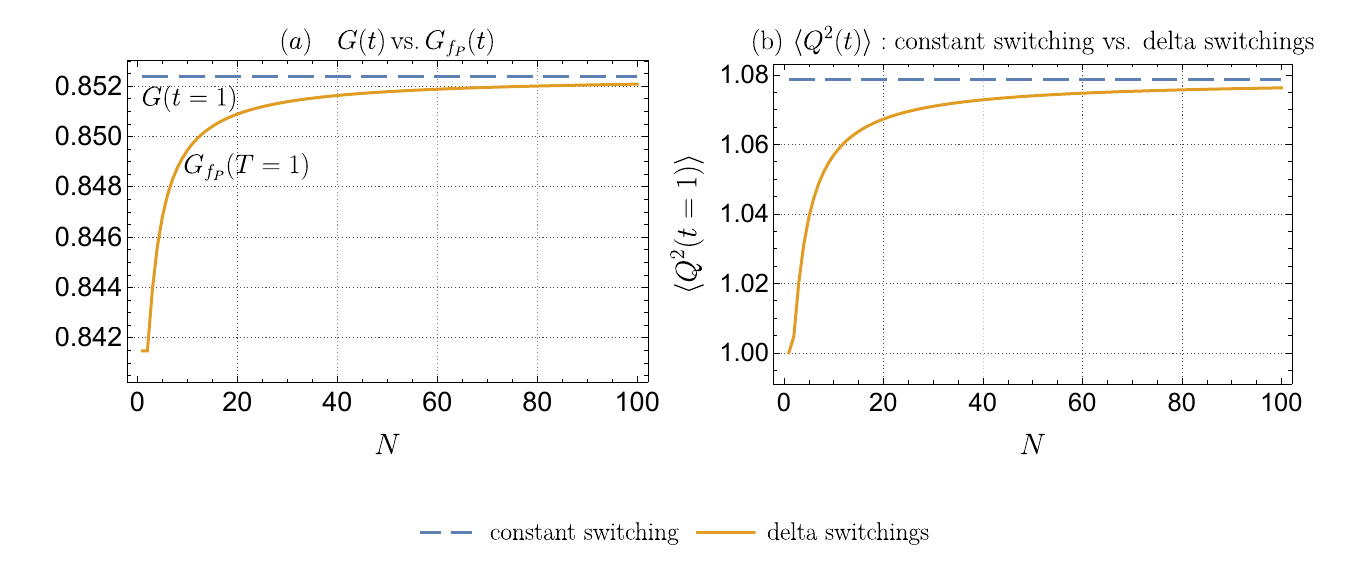}
\caption{Comparison between the solutions to the QLE with a constant switching (dashed line) and our train of delta switchings (solid curve). 
The horizontal axis is taken to be the number of delta switchings $N$. 
Here, we choose $\kappa/\Omega=0.1, \Lambda/\Omega=2$. 
(a) The function $G_{f_P}$ at fixed time $T=1$ (in units of $\Omega$) asymptotes to $G(t=1)$ as the number of delta switchings $N$ increases. 
(b) The correlation function evaluated with the train of delta switchings also asymptotes to the result for constant switching. 
}
\label{fig:FigResults}
\end{figure*}

For concreteness, let us insert $t=t_k$ into $\Gamma(t)$ in \eqref{eq:Gamma LD}: 
\begin{align}
    \Gamma_k
    &=
        \kappa \Lambda^2 
        e^{ -\Lambda T |k|/N }
        \sgn(k)\,. \label{eq:discrete Gamma data}
\end{align}
The DTFT of $\Gamma_k$ reads\footnote{Recall that we initially introduced a train of $N$ delta switchings from $k=1$ to $k=N$ [see \eqref{eq:train of delta}]. In order to employ the DTFT, we extended $k$ from $k\in \{ 1,2,\ldots, N \}$ to $k \in \mathbb Z$ by assuming that the amplitude of switching $\chi(t_k)$ is zero outside the support $t\in [0,T]$.} 
\begin{align}
    \breve \Gamma(\overline\omega)
    &=
        \sum_{k=-\infty}^\infty 
        \Gamma_k e^{\ii \overline\omega k} 
    =
        \dfrac{ -\ii \kappa \Lambda^2 \sin \overline\omega }{ \cos \overline\omega - \cosh (\Lambda T/N) }\,,
\end{align}
and we define our spectral density $s(\overline\omega)$ for the discretized data by $2\ii s(\overline\omega) \equiv \breve \Gamma(\overline\omega)$ analogous to \eqref{eq:spectral density continuous-time FT} so that 
\begin{align}
    s(\overline\omega)
    &=
        -\dfrac{1}{2} 
        \dfrac{ \kappa \Lambda^2 \sin \overline\omega }{ \cos \overline\omega - \cosh (\Lambda T/N) }\,. \label{eq:spectral density DTFT}
\end{align}
We again stress that $s(\overline\omega)$ is $2\pi$-periodic.

The inverse DTFT gives us the relation
\begin{align}
    \Gamma_k
    &=
        \dfrac{\ii }{\pi} 
        \int_{-\pi}^\pi
        \dd \overline\omega\,
        s(\overline\omega) e^{-\ii \overline\omega k}\,. \label{eq:Gamma_k in terms of s}
\end{align}
Applying the same logic used in \eqref{eq:xi correlation}, the discretized noise correlation function in the thermal state is given by 
\begin{align}
    \braket{\xi_k \xi_{k'}}
    &=
        \dfrac{1}{\pi}
        \int_{-\pi}^\pi 
        \dd \overline\omega\,
        \dfrac{s(\overline\omega)}{ e^{\beta \overline\omega}-1 }
        e^{-\ii \overline\omega (k-k')}\,. \label{eq:xi DTFT correlation}
\end{align}

Summarizing, we use the LD spectral density $\sigma(\omega)$ in \eqref{eq:LD spectral density} and the associated quantities like \eqref{eq:xi correlation} and \eqref{eq:Gamma LD} when the switching function is continuous. 
On the other hand, when we employ the train of delta-switchings, we instead use the $2\pi$-periodic spectral density $s(\overline\omega)$ in \eqref{eq:spectral density DTFT} and the related discretized data \eqref{eq:xi DTFT correlation}. 
These two spectral densities give essentially the same memory effects in the time-domain, \eqref{eq:Gamma LD} and \eqref{eq:discrete Gamma data}, respectively, and they are related by the Poisson summation formula: 
\begin{align}
    s(\omega T/N)
    &=
        \dfrac{N}{T}
        \sum_{k=-\infty}^\infty 
        \sigma(\omega - 2\pi k N/T)\,.
\end{align}

\subsection{Comparison to constant switching}\label{subsec:long interaction}
We numerically demonstrate that our solution \eqref{eq:main result} to the QLE using the delta switchings agrees with the well-known result of the continuous constant switching \eqref{eq:cont sol}.

Consider the well-known solution \eqref{eq:cont sol} to the QLE with $\chi(t)=1$. 
We choose the LD spectral density $\sigma(\omega)$ given in Eq.~\eqref{eq:LD spectral density} and evaluate $G(t)$ and $\braket{Q^2(t)}$. 
We then compare these to our solution \eqref{eq:main result} with the DTFT-version of LD spectral density $s(\omega)$ given in Eq.~\eqref{eq:spectral density DTFT}. 
Here, we choose $\chi(t_k)=1$ for all $k\in \{ 1,2,\ldots, N \}$.

Figure~\ref{fig:FigResults}(a) shows $G_{f_P}(T)$ evaluated against the total number of delta switchings $N$. 
Here, we choose $T=1$. 
As the number of switching times $N$ increases, our $G_{f_P}(T)$ asymptotes to the well-known continuous solution $G(t)$ at $t=1$. 
This indicates that the delta-switching method well-approximates the solution derived with the continuous switching. 
We also evaluate an element of covariance matrix $\braket{Q^2(t)}$ in Fig.~\ref{fig:FigResults}(b) when the system is initially prepared in a coherent state. 
As $G(t)$ is already well-approximated by $G_{f_P}(t)$, the correlation function $\braket{Q^2(t)}$ is also approximated by delta switchings when $N$ is large enough.

\section{Conclusion}\label{sec:conclusion}

By employing the train of delta-switchings method, we analytically solved a generic inhomogeneous integro-differential equation. 
The advantage of this method is that it is applicable to those with a nonstationary memory kernel, and that it allows us to visualize the memory effect using diagrams in Figs.~\ref{fig:diagrams} and \ref{fig:N2diagram}, which can also help us to construct the solution without actually solving the equation. 
We applied our solution \eqref{eq:main solution} to the Jaynes--Cummings model and the Caldeira--Leggett model. 
We found that our method well-approximates the well-known exact solutions in the continuum limit, and that the memory propagation between adjacent times corresponds to Markovian behavior.

Our method can also be applied to quantum systems in the relativistic settings (see the recent development in \cite{Sanchez.Covariant.Nonperturbative.2025}). 
In this case, the system is modeled by (a qubit-type or a harmonic oscillator-type) Unruh-DeWitt particle detector \cite{Unruh1979evaporation, DeWitt1979} coupled to a quantum field in curved spacetime. 
However, it is crucial to introduce a smearing function (i.e., a detector's size) to avoid the UV divergences due to the nature of delta-switchings. 
This is currently under investigation by the authors.

\acknowledgments{
We thank the anonymous referees for constructive comments and suggestions. 
K. G.-Y. was partially supported by Grant-in-Aid for Research Activity Start-up (Grant No. JP24K22862) and by Grant-in-Aid for JSPS Fellows (Grant No. JP25KJ0048). 
T. T. was supported by R7 (2025) Young Researchers Support Project, Faculty of Science, KYUSHU UNIVERSITY. T. T. gratefully acknowledges the support of K2SPRING for providing the resources necessary for this research.
 
}

\appendix

\section{Review of Markovian dynamical maps}\label{app:Markovian review}

In this section, we introduce the concept of Markovian dynamical maps. 
The quantum Markovianity of a quantum dynamical map is often characterized by the completely positive trace-preserving (CPTP) property, along with the phenomenologically motivated semigroup property, which ensures an irreversible dynamical process. 
These properties are equivalent to the dynamical map satisfying the well-known GKSL master equation~\cite{breuer2002theory},\footnote{The CPTP and semigroup properties are equivalent to the generalized GKSL master equation, where the Hamiltonian, coupling, and Kraus operators can depend on time.} forming the foundation of Markovian quantum dynamics.

In this work, we define Markovian dynamics as those that can be expressed as a CPTP semigroup map. 
However, we emphasize that this is not the only definition of Markovianity, as various alternative characterizations exist \cite{li2018concepts}.

Let us begin by defining Markovianity~\cite{rivas2012open}. 
Consider a dynamical map that maps a system's arbitrary quantum state from $t=0$ to time $t$, $\mathcal{E}_{t,0} : \rho\ts{s}(0) \mapsto \rho\ts{s}(t)$. 
Dynamical maps can be utilized even when a system is initially entangled with an environment. 
A \textit{completely-positive and trace-preserving (CPTP) map} is a special class of dynamical maps, where the system's input state $\rho\ts{s}(0)$ is assumed to be uncorrelated with the environment. 
For example, a map induced by $\rho\ts{s}(t)=\Tr\ts{E}[ U_t \rho\ts{tot}(0 ) U_t^\dag ]$ is a CPTP if the system is initially uncorrelated with the environment.

A Markovian dynamics can be characterized in terms of the divisibility of a CPTP map. 
Suppose a dynamical map $\mathcal E_{t,0}$ is a CPTP map and it can be decomposed into two maps: 
\begin{equation}\label{Eq: semi-group property}
    \mathcal{E}_{t,0} 
    = \mathcal{E}_{t,s}\circ \mathcal{E}_{s,0}\,, 
    \quad \forall t\geq  s \geq 0\,.
\end{equation}
In general, however, it is not necessarily true that both $\mathcal{E}_{t,s}$ and $\mathcal{E}_{s,0}$ are CPTP maps. 
For example, $\mathcal{E}_{s,0}$ can be a CPTP map, but $\mathcal{E}_{t,s}$ is generally not, as the first CPTP map $\mathcal{E}_{s,0}$ entangles the system and the environment. 
A CPTP map $\mathcal{E}_{t,0}$ is called a \textit{Markovian dynamical map} if \textit{both} dynamical maps $\mathcal{E}_{t,s}$ and $\mathcal{E}_{s,0}$ are CPTP maps. 
This is also known as the one-parameter semigroup property, and it can be shown that it is equivalent to the generalized GKSL equation~\cite{rivas2012open}.

\section{Proof of Markovianity in the damped Jaynes–Cummings Model
}\label{sec:Markov JC}

As discussed in Sec.~\ref{subsec:nonmarkov JC model}, the semigroup property of the quantum channel reduces to a property of the time-evolution function of the probability amplitude, given by Eq.~(\ref{eq:JC pure delta sol}), as follows: 
\begin{align}
    \mathcal T(t)
    &=
        \mathcal T(t,t')\mathcal T(t')\,,\label{Eq: semigroup alpha}
\end{align}
for any $t' \in [0,t)$. 
Below, we prove that our solution \eqref{eq:JC pure delta sol} satisfies this Markovianity condition if the memory propagation in Fig.~\ref{fig:N2diagram} is restricted to nearest-neighbor propagation $t_l \xrightarrow{\Sigma} t_{l+1}$, where $t_l\equiv lT/N$. 
In practice, we impose the Markov approximation $\Lambda T/N\gg 1$ to ensure that the nearest-neighbor propagation is dominant. 
To this end, consider $\mathcal T(t)$ in \eqref{eq:JC pure delta sol} only composed of the neighboring propagation $\Sigma(t_{l+1}, t_{l})$, which reads 
\begin{align}
    \mathcal T(t,0)
    &=
        1 
        + 
        \sum_{i_1=1}^{N-1} X_{i_1}
        +
        \sum_{i_1=1}^{N-1} 
        \sum_{i_2=1}^{i_1 - 1} X_{i_1} X_{i_2}
        + \cdots \notag \\
    &=
        \mathsf T 
        \exp 
        \kako{
            \sum_{j=1}^{N-1} X_j
        },
        \label{Eq: markov time evolution T}
\end{align}
where 
\begin{align}
    &X_j
    \equiv 
        -\dfrac{T^2}{N^2}
        \Sigma(t_{j+1}, t_j)\,,
\end{align}
and $\mathsf T$ denotes the time-ordering symbol. 
Note that we take $t=t_N$. 
Based on the above expression, the time-evolution function $\mathcal T(t_b, t_a)$ from time $t_a$ to $t_b$, with integers $a<b(<N)$, can be written as 
\begin{align*}
    \mathcal T (t_b,t_a)
    =
    \mathsf T 
        \exp 
        \kako{
             \sum_{j=a}^{b-1} X_j
        }\,.
\end{align*}

We now proceed to verify the semigroup property \eqref{Eq: semigroup alpha}.
Since $t>t'>0$,  without loss of generality, let us denote $t'=t_a$ for any integer $a\in (1,N)$, and $t=t_N$.
Then we find that 
\begin{align}
    \mathcal T(t,t_a) \mathcal T(t_a)
    &=
        \mathsf T 
        \exp 
        \kako{
            \sum_{j=a}^{N-1} X_j
        }
        \mathsf T 
        \exp 
        \kako{
            \sum_{j=1}^{a-1} X_j
        } \notag \\
    &=
        \mathsf T 
        \exp 
        \kako{
            \sum_{j=1}^{N-1} X_j
        } 
    =
        \mathcal T(t)\,,
\end{align}
which completes our proof.

\section{Measures of non-Markovianity}\label{app:nonMarkov measure}

In this appendix, we show that when $ \kappa T/N$ is finite, the $g$-function (defined below) used in the RHP measure equals the decay rate plus a correction term. This correction term vanishes in the limit $\kappa T/N \rightarrow 0$, rendering a continuous switching case.

First, let us state the RHP measure introduced in Ref.~\cite{Rivas.nonMarkovMeasure.2010}. Let  $\mathcal{E}_{t,0} : \mathcal{L}(\mathcal{H})\rightarrow \mathcal{L}(\mathcal{H})$ be a Markovian dynamical map. Let us consider an infinitesimal evolution beyond the final time $t$ with $\mathcal{E}_{t+\epsilon , t}$. Then a composite map is given by  
\begin{align}
    \mathcal{E}_{t+\epsilon ,0} 
    = \mathcal{E}_{t+\epsilon ,t} \circ \mathcal{E}_{t,0}.
\end{align}
If the infinitesimal increment $\mathcal{E}_{t+\epsilon ,t}$ is a CPTP map, then $\mathcal{E}_{t+\epsilon , 0}$ is also a CPTP map. 
This small incremental evolution can be done many times to construct any time evolution beyond $t$, which ensures the semi-group property: 
\begin{align}
    \mathcal{E}_{t_1,t_2} = \mathcal{E}_{t_1 , s } \circ \mathcal{E}_{s , t_2},
\end{align}
where $t_1 \geq s \geq t_2 \geq 0$. 

The CPTP property of the dynamical map $\mathcal{E}_{t+\epsilon,t}$ is ensured by showing that the Choi matrix of the map $\mathcal{E}_{t+\epsilon , t}$ is positive definite, which is true if the trace-norm of the Choi matrix is equal to 1:
\begin{subequations}
\begin{align}
    \|(\mathcal{E}_{t+\epsilon , t}\otimes \id\ts{A}) \left[\Psi^+\ts{SA}\right]\| =1 & \text{ iff }\mathcal{E}_{t+\epsilon , t}\text{ is CP,}\\
     \|(\mathcal{E}_{t+\epsilon , t}\otimes \id\ts{A} )\left[\Psi^+\ts{SA}\right]\| >1 & \text{ iff }\mathcal{E}_{t+\epsilon , t}\text{ is non-CP,}
\end{align}
\end{subequations}
where `S' and `A' refer to the system and ancilla, respectively, and $\Psi^+\ts{SA}\equiv (\ket{01}+\ket{10})(\bra{01}+\bra{10})/2$ is a normalized maximally entangled state. 
Note that taking a different Bell basis would not change the eigenvalues of the Choi matrix.

Using the above relation authors in Ref.~\cite{Rivas.nonMarkovMeasure.2010} constructed the $g$-function:
\begin{align}
    g(t) 
    \coloneqq 
        \lim_{\epsilon \rightarrow 0^+} 
        \dfrac{ \|(\mathcal{E}_{t+\epsilon , t}\otimes \id\ts{A}) \left[\Psi^+\ts{SA}\right]\| -1}{\epsilon} ,
\end{align}
which gives zero if and only if the map $\mathcal{E}_{t+\epsilon , t}$ is CP, and becomes positive if and only if $\mathcal{E}_{t+\epsilon , t}$ is non-CP. Integrating this function over all time scales, we obtain the RHP measure:
\begin{align}
    \mathcal{I} 
    \coloneqq 
        \int_0^\infty \dd t\, g(t),
\end{align}
which can be used as a measure of Markovianity: $\mathcal{I}=0$ indicating the Markovianity, and $\mathcal{I} \neq 0$ indicating the non-Markovianity. 
In the damped Jaynes-Cumming model the $g$-function is related to the decay rate $\gamma(t)$ of the system in the following way 
\begin{align}
    g(t) = \left\{\begin{array}{cc}
        0 & \text{ if } \gamma(t)\geq 0  \\
        -\gamma(t) & \text{ if } \gamma(t)< 0
    \end{array}\right. \,.
\end{align}
We will show below that this relation only holds in our Dirac-delta switching method when $\kappa T /N$ tends to zero.   

To find the relation between the $g$-function and the decay rate $\gamma(t)$ in our delta-switching scenario, we first need to obtain the time-convolutionless master equation for our setup. Following the method in Ref.~\cite{breuer2002theory}, we take the time derivative of the system density matrix at an arbitrary time $t$, given in Eq.~(\ref{Eq: quantum channel for JC model}):
\begin{widetext}
\begin{align}
    \dfrac{\rho\ts{s}(t+\delta)-\rho\ts{s}(t)}{\delta}
    &=
    \dfrac{1}{\delta}
    \Bigg(
    \begin{bmatrix}
        |\alpha_1(t+\delta)|^2&\alpha_0^*\alpha_1(t+\delta)\\
        \alpha_0\alpha_1^*(t+\delta) & 1-|\alpha_1(t+\delta)|^2
    \end{bmatrix}
    -
    \begin{bmatrix}
        |\alpha_1(t)|^2&\alpha_0^*\alpha_1(t)\\
        \alpha_0\alpha_1^*(t) & 1-|\alpha_1(t)|^2
    \end{bmatrix}
    \Bigg) \notag\\
    &=
    \begin{bmatrix}
        2\dot \alpha_1(t) \alpha_1^*(t)&\alpha_0^*\dot \alpha_1(t) \\
        \alpha_0\dot \alpha_1^*(t)&-2\dot \alpha_1(t) \alpha_1^*(t)
    \end{bmatrix}
    +
    \begin{bmatrix}
        \delta|\dot \alpha_1(t)|^2&0 \\
        0&-\delta|\dot \alpha_1(t)|^2
    \end{bmatrix}\,,
\end{align}
\end{widetext}
where $\delta\equiv T/N$ is the time interval between Dirac-delta switching. We have also defined a quantity $\dot \alpha_1(t)\equiv( \alpha_1(t+\delta)-\alpha_1(t)) / \delta $. Rewriting this equation in terms of the Pauli matrices $\sigma_\pm  = (\sigma_x \pm \ii \sigma_y) $, we find 
\begin{align}
    \dfrac{\rho\ts{s}(t+\delta)-\rho\ts{s}(t)}{\delta}
    =&
    \gamma(t)
    \Big[
    \sigma_+\rho\ts{s}(t)\sigma_-
    -\dfrac{1}{2}
    \{\sigma_-\sigma_+,\rho\ts{s}(t)\}
    \Big]\notag\\
    &+
    \delta h(t)
    \Tr[\rho\ts{s}(t)][\sigma_-,\sigma_+] \notag\\
    \equiv&
    \mathcal L_{t}[\rho\ts{s}(t)]\,,
\end{align}
where we have defined the two functions
\begin{align}
    \gamma(t) \equiv -2 \text{Re} \left(\frac{\dot{\alpha_1}(t)}{\alpha_1(t)}\right) , \quad h(t) \equiv \left|\dot{\alpha_1}(t)\right|^2,
\end{align}
where $\gamma(t) $ corresponds to the decay rate of our system. Notice that if we take $\delta|\dot \alpha_1(t)|^2 \rightarrow 0$ in the above equation, we obtain the usual time-convolutionless master equation for the damped Jaynes-Cummings model Ref.~\cite{breuer2002theory}. 

Rewriting the left-hand side of the above equation with $\rho\ts{s}(t+\delta)=\mathcal E_{t+\delta,t}[\rho\ts{s}(t)]$, we find 
\begin{align}
    \mathcal E_{t+\delta,t}[\rho\ts{s}(t)]
    =
    (\id\ts{S}+ \mathcal L_{t} \delta)[\rho\ts{s}(t)]\,.
\end{align}
Finally, the Choi matrix of the above quantum channel gives a $4\times 4$ matrix of the following form:
\begin{align}
   &\id\ts{SA}+ 
   \delta(\mathcal L_{t} \otimes\id\ts{A})
    )[\Psi\ts{SA}^+]\notag\\
    &=\dfrac{1}{2}
    \begin{bmatrix}
        0&\delta^2 h(t)&0&0 \\
        0&1-\delta\gamma(t)+\delta^2 h(t)&1-\delta\frac{\gamma(t)}{2}&0 \\
        0&1-\delta\frac{\gamma(t)}{2}&1&-\delta^2 h(t) \\
        0&0&0&\delta\gamma(t)-\delta^2 h(t)
    \end{bmatrix}\,,
\end{align}
with the following four eigenvalues: 
\begin{align}
    &\lambda_0 =0 ,~ \lambda_1 = \frac{\delta}{2}(\gamma(t)-\delta h(t)) , \\
    & \lambda_\pm = \frac{2-\delta(\gamma(t)-\delta h(t))}{4} \notag \\
    &~\pm \frac{\sqrt{4 - 4 \delta \gamma(t) +2 \delta^2 \gamma(t)^2 - 2 \delta^3 h(t)\gamma(t)+\delta^4 h(t)^2}}{4}.
\end{align}
To calculate the trace-norm, we take the absolute value sum of all four eigenvalues. However, since our interest lies in the small time interval $T/N$, we can use the perturbation theory. By inspection, notice that eigenvalues depend on the combination $\delta^2 h(t) = \mathcal{O} (\kappa T/N)$, where $\kappa$ is the coupling strength between the system and the environment. Therefore, we can Taylor expand the trace-norm with respect to the dimensionless parameter $\kappa T/N$: 
\begin{align}
    &\|
    (\id\ts{SA}+ \delta
    (\mathcal L_{t} \otimes\id\ts{A})
    )[\Psi\ts{SA}^+
    ]
    \|\notag\\
    &\quad= 
    \left|1-\dfrac{1}{2}\delta\gamma(t)\right|
    +\dfrac{\delta}{2}|\gamma(t)|
    +\mathcal O((\kappa \delta)^2) \notag\\
    &\quad=
    1-\dfrac{1}{2}\delta\gamma(t)+\dfrac{\delta}{2}|\gamma(t)|+\mathcal O((\kappa \delta)^2),
\end{align}
where we observe that there is a correction term of order $(\kappa \delta)^2 = (\kappa T / N)^2$. Calculating the $g$-function, we find 
\begin{align}
    g(t) = 
    \begin{cases}
    0 +\mathcal O(\kappa^2 \delta)& \text{if}\quad \gamma(t)\ge 0 \\
    -\gamma(t) +\mathcal O(\kappa^2 \delta) & \text{if}\quad \gamma(t)< 0
    \end{cases}.
\end{align}    
Since this is a small correction term, we take the decay rate $\gamma(t)$ to calculate the RHP measure in Sec.~\ref{sec:JC model}. However, we note that in the strong coupling limit or the long-time interval case, the correction term $\kappa T/N$ cannot be ignored, and the decay rate is no longer the measure of Markovianity. It is interesting to study such cases, but since this is outside the scope of this paper, we leave it for future work.


\section{Markovianity and the covariance matrix} \label{app:Markov}

\subsection{Transformation of covariance matrices}

As described in Sec.~\ref{subsec:correlation function and CM}, Gaussian states are fully determined by the first moment $\bar{\bm R}\equiv \braket{\bm R}$ and the covariance matrix $\mathbb{V}$. 
Any tensor product of Gaussians is Gaussian, and Gaussian operations are those CP maps preserving the Gaussianity. 
In particular, the full Hamiltonian in the Caldeira-Leggett model generates a Gaussian unitary. 
If the initial joint state is a product of Gaussian states, $\rho\ts{tot}(0)=\rho\ts{s}(0) \otimes \rho\ts{E}(0)$, then the final reduced state of the system, $\rho\ts{s}(t)=\Tr\ts{E}[ U_t \rho\ts{tot}(0) U_t^\dag ]$, after the Gaussian unitary time-evolution $U_t \equiv U(t,0)$ remains Gaussian.

Since Gaussian states are characterized by covariance matrices,  
we can equivalently track the evolution of $\mathbb V(t)$. 
In phase space in which covariance matrices are defined, tensor products $\otimes$ become direct sums $\oplus$, and Gaussian unitaries correspond to symplectic matrices.
Therefore, computing the system's final density matrix $\rho\ts{s}(t)=\Tr\ts{E}[ U_t \rho\ts{tot}(0 ) U_t^\dag ]$ is equivalent to computing the system's covariance matrix $\mathbb V(t)$ by performing a partial trace on the final joint covariance matrix $\mathbb S_t (\mathbb V(0) \oplus \mathbb V\ts{E}(0) ) \mathbb S_t^\intercal $, where $\mathbb S_t$ is the symplectic matrix corresponding to the Gaussian unitary operator $U_t$ and $\mathbb V\ts{E}(0)$ is the environment's initial covariance matrix.

It is well-known that if the joint state is initially \textit{separable} Gaussian, then the system's covariance matrix $\mathbb{V}$ evolves as \cite{serafini.QCV}
\begin{align}
    \mathbb{V}(t)
    &=
        \mathbb T_t \mathbb V(0) \mathbb T^\intercal_t 
        + 
        \mathbb N_t\,, \label{eq:CM evolution}
\end{align}
where $\mathbb T_t \equiv \mathbb{T}(t,0)$ and $\mathbb N_t \equiv \mathbb{N}(t,0)$ are $2\times 2$ matrices that transform the system's covariance matrix at $t=0$ to time $t$. 
However, if the initial joint state is not separable, no such simple affine form in terms of $\mathbb V(0)$ alone exists.

As an example, suppose we apply two Gaussian unitaries $U_t\equiv U(t,0)$ and $U_{T-t}\equiv U(T,t)$ to a separable initial joint Gaussian state $\rho\ts{tot}(0)$. 
The system's final state, $\rho\ts{s}(T)=\Tr\ts{E}[ U_{T-t} U_t \rho\ts{tot}(0) U^\dag_t U^\dag_{T-t} ]$, is Gaussian, and the corresponding covariance matrix of the system can be expressed as 
\begin{align}
    \mathbb{V}(T)
    &=
        \mathbb T_T \mathbb V(0) \mathbb T^\intercal_T
        + 
        \mathbb N_T\,, \notag
\end{align}
where $\mathbb T_T$ and $\mathbb N_T$ are square matrices corresponding to the \textit{entire} Gaussian time-evolution $U_{T-t} U_t (\equiv U_T)$. 
Similarly, for the first evolution $U_t$ alone, 
\begin{align}
    \mathbb V(t)
    &=
        \mathbb T_t \mathbb V(0) \mathbb T^\intercal_t
        + 
        \mathbb N_t\,. \notag
\end{align}
Here, $\mathbb T_t \equiv \mathbb T(t,0)$ and $\mathbb N_t \equiv \mathbb N(t,0)$ with $t\in (0,T)$ correspond to the time-evolution under $U_t$. 
However, the following relation generally does not hold: 
\begin{align}
    \mathbb V(T)
    &=
        \mathbb T_{T-t} \mathbb V(t) \mathbb T_{T-t}^\intercal 
        + 
        \mathbb N_{T-t}\,, \label{eq:transform V(t) to T}
\end{align}
where $\mathbb T_{T-t}\equiv \mathbb T(T,t)$ and $\mathbb N_{T-t}\equiv \mathbb N(T,t)$ are meant to transform the covariance matrix $\mathbb V(t)$ to $\mathbb V(T)$. 
This is because the system entangles with the environment after the first Gaussian unitary $U_t$, thereby the actual expression that relates $\mathbb V(t)$ to $\mathbb V(T)$ is more complicated. 
This fact is relevant in our Dirac-delta method as the entire Gaussian unitary operator can be decomposed into a product $U_T= U_{T-t_{N-1}} U_{t_{N-1}-t_{N-2}}\ldots U_{t_1}$, where each unitary operator is generated by a delta-coupling.

\subsection{Transformation of covariance matrices under the Markovian dynamics}




Next, we consider how the above Markovianity condition~\eqref{Eq: semi-group property} applies to the formulation using covariance matrices in Gaussian states. 
In particular, we show that if the time-evolution is described by a Markovian dynamical map, then the matrices $\mathbb T$ and $\mathbb N$ satisfy 
\begin{align}\label{Eq: Markovianity condtion}
    \mathbb{T}_{t} = \mathbb{T}_{t-s}\mathbb{T}_{s} \,, \quad \mathbb{N}_{t} = \mathbb{T}_{t-s} \mathbb{N}_{s}\mathbb{T}^\intercal_{t-s} 
    + \mathbb{N}_{t-s}\,,
\end{align}
where $\mathbb T_{t-s}\equiv \mathbb T(t,s)$ and $\mathbb N_{t-s}\equiv \mathbb N(t,s)$ are the matrices associated with a CPTP map $\mathcal E_{t,s}$. 



To show this, consider the system's initial Gaussian state $\rho\ts{s}(0) \equiv \rho\ts{s}[ \bar{\bm R}(0), \mathbb V(0) ]$. 
Assuming the joint initial state is a Gaussian product state and that the entire system evolves under a Gaussian unitary, the system's state at time $t$ is expressed as 
\begin{align}
    \mathcal E_{t,0}[ \rho\ts{s}[\bar{\bm R}(0), \mathbb V(0)] ]
    =
        \rho\ts{s}[ \mathbb T_t \bar{\bm R}(0), \mathbb T_t \mathbb V(0) \mathbb T_t^\intercal + \mathbb N_t ]\,,
\end{align}
where $\mathcal E_{t,0}$ is a CPTP map and we used the relation \eqref{eq:R and V under UDM}. 
Suppose the CPTP map $\mathcal{E}_{t,0}$ is a Markovian dynamical map. 
Then, we have the following relations: 
\begin{align}
    \rho\ts{s}[ \bar{\bm R}(s), \mathbb V(s) ]
    &=
        \mathcal E_{s,0}[ \rho\ts{s}[\bar{\bm R}(0), \mathbb V(0)] ] \notag \\
    &=
        \rho\ts{s}[ \mathbb T_s \bar{\bm R}(0), \mathbb T_s \mathbb V(0) \mathbb T_s^\intercal + \mathbb N_s ]\,, \notag \\
    \mathcal E_{t,s}[ \rho\ts{s}[\bar{\bm R}(s), \mathbb V(s)] ]
    &=
        \rho\ts{s}[ \mathbb T_{t-s} \bar{\bm R}(s), \mathbb T_{t-s} \mathbb V(s) \mathbb T_{t-s}^\intercal + \mathbb N_{t-s} ]\,. \notag
\end{align}
Thus, the semigroup property $\mathcal{E}_{t,0}=\mathcal{E}_{t,s}\circ \mathcal{E}_{s,0}$ leads to \eqref{Eq: Markovianity condtion}, which completes our proof.


\section{Proof of Markovianity in Caldeira-Leggett model}\label{app:BornMarkov}

In Appendix~\ref{sec:Markov JC}, we showed that the Markovianity is realized when the memory kernel $\Sigma(t,t')$ is restricted to nearest-neighbor propagation. 
In this section, we present an analogous analysis in the Caldeira–Leggett model.

The explicit forms of the transformation matrices $\mathbb{T}_t$ and $\mathbb{N}_t$ are given by
\begin{subequations}
    \begin{align}
    \mathbb{T}_{t} 
        &
        =
        \begin{bmatrix}
        G_{f_Q}(t) & G_{f_P}(t) \\
        \dot G_{f_Q}(t) & \dot G_{f_P}(t)
        \end{bmatrix}\,, \\
    \mathbb{N}_t 
        &
        =
        \begin{bmatrix}
        N_{QQ}(t) & N_{QP}(t) \\
        N_{PQ}(t) & N_{PP}(t)
        \end{bmatrix}\,,
\end{align}\label{Eq: CL T N func}
\end{subequations} 
where the functions $G_{f_Q}(t)$ and $G_{f_P}(t)$ are defined in (\ref{eq:main result functions}), and the matrix components $N_{QQ}(t)$, $N_{QP}(t)$, $N_{PQ}(t)$, and $N_{PP}(t)$ are specified in \eqref{eq:Noises for delta}. 

Next, 
we consider transformation matrices \eqref{Eq: CL T N func} composed only of the neighboring propagation $\Sigma(t_{l+1},t_l)$. 
By utilizing the trigonometric identities, 
\begin{subequations}
    \begin{align}
        \sin [\Omega (t-t_{j})] 
        &=
        \sin(\Omega t) \cos(\Omega t_{j}) - \cos(\Omega t) \sin(\Omega t_{j})\,,\\
        \cos [\Omega (t-t_{j})] 
        &=
        \cos(\Omega t) \cos(\Omega t_{j}) + \sin(\Omega t) \sin(\Omega t_{j})\,,
    \end{align}
    \label{Eq: trigonometric identities}
\end{subequations}
we find 
\begin{subequations}
\begin{align}
    \mathbb{T}^{(\mathrm M)}_{t}
    &=
    \bm F_t
    +
    \bm F_t
    \sum_{i_1 =1}^{N-1}
    \bm P_{i_1}
    +
    \bm F_t
    \sum_{i_1 =1}^{N-1}
    \sum_{i_2 =1}^{i_1-1}
    \bm P_{i_1}
    \bm P_{i_2}
    +\cdots\notag\\
    &=
    \bm F_t
    \,
    \mathsf T 
        \exp 
        \kako{
            \sum_{j=1}^{N-1} \bm P_j
        }\,,
        \\
    \mathbb{N}^{(\mathrm M)}_{t}
    &=
    \sum_{k,k'=1}^N
        \bm F_t
    \,
    \mathsf T 
        \exp 
        \kako{
            \sum_{j=1}^{N-1} \bm P_j
        }
        \bm g(t_k)
        \nu_{k,k'} \notag\\
    &\qquad
    \times
        \kagikako{
        \bm F_t
    \,
    \mathsf T 
        \exp 
        \kako{
            \sum_{j=1}^{N-1} \bm P_j
        }
        \bm g(t_{k'})
        }^\intercal
        \,.
\end{align}
\label{Eq:CL T N markov mat}
\end{subequations}
Here, $\bm F_{t,t'} \coloneqq \bm F(t,t')$ is a $2\times 2$ matrix 
\begin{align*}
    \bm F_{t,t'}
     &=
    \begin{bmatrix}
        \cos[\Omega (t-t')]& \sin[\Omega (t-t')]/\Omega \\
        -\Omega \sin[\Omega (t-t')]&\cos[\Omega (t-t')]
    \end{bmatrix}\,,
\end{align*}
representing the unitary evolution of the system, $\bm g (t)=[-\sin(\Omega t)/\Omega, \cos(\Omega t)]^\intercal $ is a vector constructed from free Green's function for a quantum harmonic oscillator, and 
\begin{widetext}
\begin{align}
    \bm P_j
    &=
    \begin{bmatrix}
        -\sin(\Omega t_{j+1}) X_j \cos(\Omega t_j )/\Omega &  -\sin(\Omega t_{j+1}) X_j \sin(\Omega t_j )/\Omega^2 \\
       \cos(\Omega t_{j+1}) X_j \cos(\Omega t_j )&\cos(\Omega t_{j+1}) X_j \sin(\Omega t_j )/\Omega
    \end{bmatrix}
    \,,
    \end{align}
\end{widetext}
with $X_j=-\Sigma(t_{j+1},t_j) T^2 /N^2$.

For any integer $a\in (1,N)$ and $t=t_N$, the Markovianity condition \eqref{Eq: Markovianity condtion} can be shown as 
\begin{align}
    \mathbb{T}_{t,t_a}\mathbb{T}_{t_a}
    &=
    \bm F_{t,t_a} \bm F_{t_a}
    \mathsf T 
        \exp 
        \kako{
             \sum_{j=a}^{N-1} \bm P_j
        }\,
         \mathsf T 
        \exp 
        \kako{
             \sum_{j=1}^{a-1} \bm P_j
        }\,\notag\\
    &=
    \bm F_t
    \,
    \mathsf T 
        \exp 
        \kako{
            \sum_{j=1}^{N-1} \bm P_j
        }
    =\mathbb{T}_{t} \,,
\end{align}
where we used the trigonometric identities \eqref{Eq: trigonometric identities} in the second equality. 
From a similar calculation, one can show that the noise matrix $\mathbb N_t^{(\mathrm M)}$ satisfies the composition law \eqref{Eq: Markovianity condtion}.

\bibliography{ref}

\end{document}